%% file: main.tex
\documentclass[10pt,twocolumn]{article}
\usepackage[margin=0.72in]{geometry}
\usepackage{amsmath,amssymb,booktabs,graphicx,microtype,url}
\usepackage{tikz,pgfplots}
\usepackage{placeins}
\usepackage{needspace}
\usetikzlibrary{arrows.meta,positioning,fit,calc,backgrounds}
\usepgfplotslibrary{groupplots}
\pgfplotsset{compat=1.18}
\usepackage[hidelinks]{hyperref}
\title{One Global Beam Across Many GPUs: High-Throughput Beam Search at Billion-Record Frontier Scale}
\author{Ivan Litvak\\Independent Researcher}
\date{}

\newcommand{\beam}{B}
\newcommand{\frontier}{\mathcal{F}}
\newcommand{\cand}{\mathcal{C}}

\begin{document}
\maketitle

\begin{abstract}
Beam search repeatedly makes many children, removes duplicates, and keeps the
best $B$.  We show how many GPUs can perform these steps as one search even
when the retained set does not fit on one device.  States and candidates stay
on the GPUs; the CPU receives only small control and ancestry records.  We
prove that an abstract distributed pipeline returns the monolithic reduced-key
top-$B$ for the same complete candidates, integer scores, Hash128 equivalence,
and a fully specified physical-layout tie order, provided that its implementation performs one
global reduction per key and race-free routing.  A static audit of the
historical one-T4 pin leaves cross-buffer uniqueness unresolved and identifies
a separate multi-rank scatter risk; neither is a reproduced failure, and other
revisions inherit neither defects nor correctness without comparison.  On eight
H200 GPUs, a Cube4 run completed saturated depth 8 at
$B_{\rm eff}=2{,}900{,}361{,}216$ in $931.266$ s: $69{,}608{,}669{,}184$
nominal parent--generator pairs, or a derived $74.746$ million pairs/s.    A separate
two-T4 Megaminx run yielded a derived $30.274$ million nominal
parent--generator pairs/s at
$B_{\rm eff}=82{,}837{,}504$.  Those tasks and hardware differ, so they do not
establish strong scaling.  Separately, on one eight-RTX-3060 host, fixed-count
eight-GPU speedup was $7.487$ under a common execution profile and $5.605$
under selected stable profiles; weak actual-work throughput gain was $6.113$
and $7.295$.  These profile-sensitive ratios use each series' own one-GPU
baseline.
\end{abstract}

\section{Introduction and Problem}
\begin{figure}[!htbp]
\centering
\begin{tikzpicture}[
  >=Latex,
  state/.style={circle,draw,minimum size=7mm,font=\small\bfseries},
  scorelabel/.style={font=\scriptsize},
  flow/.style={->,thick,gray!75}
]
\node[font=\small\bfseries] at (2.5,6.0) {parents};
\node[state,fill=blue!15] (p1) at (1.5,5.4) {$P_1$};
\node[state,fill=blue!15] (p2) at (3.5,5.4) {$P_2$};
\node[font=\small\bfseries,align=center] at (5.45,4.75)
  {generate\\and score};
\node[state,fill=yellow!25] (a1) at (0.55,3.95) {$A$};
\node[state,fill=yellow!25] (b4) at (1.85,3.95) {$B$};
\node[state,fill=yellow!25] (b2) at (3.15,3.95) {$B$};
\node[state,fill=yellow!25] (c3) at (4.45,3.95) {$C$};
\foreach \u/\v in {p1/a1,p1/b4,p2/b2,p2/c3} \draw[flow] (\u)--(\v);
\node[scorelabel,below=1mm of a1] {$s=1$};
\node[scorelabel,below=1mm of b4] {$s=4$};
\node[scorelabel,below=1mm of b2] {$s=2$};
\node[scorelabel,below=1mm of c3] {$s=3$};
\draw[->,very thick] (2.5,3.35)--(2.5,2.85);
\node[font=\small\bfseries] at (2.5,2.65) {one record per key};
\node[state,fill=green!20] (ua) at (1.05,1.95) {$A$};
\node[state,fill=green!20] (ub) at (2.5,1.95) {$B$};
\node[state,fill=green!20] (uc) at (3.95,1.95) {$C$};
\node[scorelabel,below=1mm of ua] {best $s=1$};
\node[scorelabel,below=1mm of ub] {best $s=2$};
\node[scorelabel,below=1mm of uc] {best $s=3$};
\node[fit=(ua)(ub)(uc),inner sep=1mm] (reducedset) {};
\node[font=\small\bfseries] at (2.5,0.65) {next frontier};
\node[draw,rounded corners,fill=red!12,minimum width=3.4cm,
      minimum height=1.15cm] (frontbox) at (2.5,-0.15) {};
\node[state,fill=white] (fa) at (1.75,-0.15) {$A$};
\node[state,fill=white] (fb) at (3.25,-0.15) {$B$};
\node[scorelabel,below=1mm of fa] {$s=1$};
\node[scorelabel,below=1mm of fb] {$s=2$};
\draw[->,very thick] (reducedset.east)--++(4mm,0)|-(frontbox.east);
\end{tikzpicture}
\caption{Beam search in one picture. Each parent produces children; records
with the same identity key retain their best score; the best $B$ distinct-key
records become the next frontier. Smaller scores are better in this paper.}
\label{fig:beam-basics}
\end{figure}
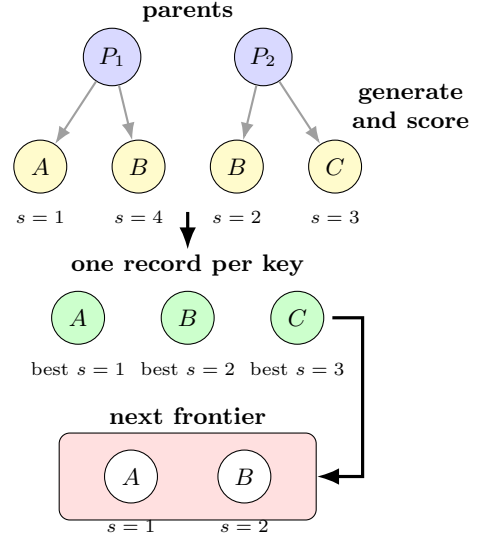

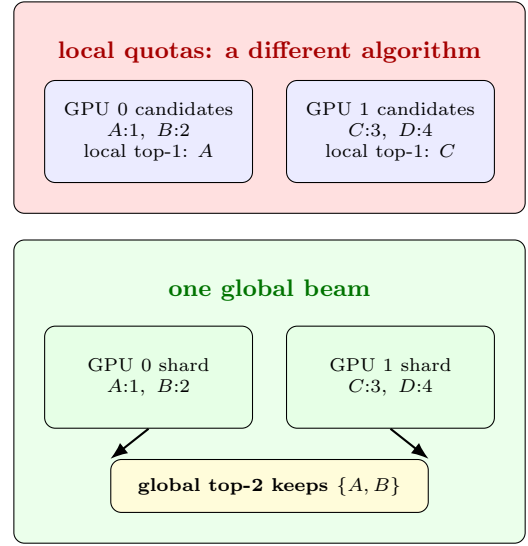
\begin{figure}[!htbp]
\centering
\begin{tikzpicture}[
  >=Latex,
  gpu/.style={draw,rounded corners,minimum width=2.75cm,minimum height=1.35cm,
              align=center,font=\scriptsize},
  title/.style={font=\small\bfseries}
]
\node[title,red!65!black] (ltitle) at (3.15,5.0)
  {local quotas: a different algorithm};
\node[gpu,fill=blue!8] (l0) at (1.55,3.95)
  {GPU 0 candidates\\$A{:}1,\ B{:}2$\\local top-1: $A$};
\node[gpu,fill=blue!8] (l1) at (4.75,3.95)
  {GPU 1 candidates\\$C{:}3,\ D{:}4$\\local top-1: $C$};
\begin{scope}[on background layer]
  \node[draw,rounded corners,fill=red!12,fit=(ltitle)(l0)(l1),inner sep=4mm] {};
\end{scope}
\node[title,green!45!black] (gtitle) at (3.15,1.85) {one global beam};
\node[gpu,fill=green!10] (g0) at (1.55,0.70)
  {GPU 0 shard\\$A{:}1,\ B{:}2$};
\node[gpu,fill=green!10] (g1) at (4.75,0.70)
  {GPU 1 shard\\$C{:}3,\ D{:}4$};
\node[draw,rounded corners,fill=yellow!18,below=4mm of $(g0.south)!0.5!(g1.south)$,
      minimum width=4.2cm,minimum height=7mm,font=\scriptsize\bfseries] (cut)
      {global top-2 keeps $\{A,B\}$};
\draw[->,thick] (g0.south)--(cut.north west);
\draw[->,thick] (g1.south)--(cut.north east);
\begin{scope}[on background layer]
  \node[draw,rounded corners,fill=green!8,fit=(gtitle)(g0)(g1)(cut),inner sep=4mm] {};
\end{scope}
\end{tikzpicture}
\caption{Why sharding is not the same as running independent beams. With
local quotas, a GPU can discard a state that belongs to the global top-$B$.
This work applies one global cut across all shards.}
\label{fig:global-vs-local}
\end{figure}

Beam search is easy to describe.  Start with a set of states.  Make their
children, score them, remove repeated states, and retain the best $B$:
\begin{equation}
\boxed{\text{parents}\to\text{children}\to\text{one per key}\to\text{best }B}
\label{eq:pipeline}
\end{equation}
Figure~\ref{fig:beam-basics} makes the key-reduction step explicit before the
global cut.
The retained set is the \emph{frontier}.  If it outgrows one GPU, simply giving
each GPU an independent beam changes the algorithm.  Our goal is different:
split one frontier across GPUs while preserving the result of one monolithic
beam.  Figure~\ref{fig:global-vs-local} shows why independent local quotas do
not implement that objective.

Formally, states are vertices of an implicit graph $G=(V,E)$.  Applying move
$a\in\{1,\ldots,m\}$ to state $x$ produces $A_a(x)$.  A deterministic oracle
$\sigma(x,a)$ scores that choice.  It may be learned or hand-engineered; our
experiments use neural scorers.  The implementation compares only the
quantized integer score $s(x,a)$ from Eq.~\eqref{eq:score}.  The complete depth
transformation is
\begin{equation}
 \frontier_{d-1}\xrightarrow{\times m}\cand_d
 \xrightarrow{\operatorname{Reduce}_z}U_d
 \xrightarrow{\operatorname{Top}_{\beam}}\frontier_d.
 \label{eq:depth-map}
\end{equation}
Its three stages are defined by
\begin{align}
  \cand_d &= \{(x,a,A_a(x)):x\in\frontier_{d-1},\ a\in[m]\},\nonumber\\
  U_d &= \operatorname{Reduce}_{z}
  (\cand_d;\,\operatorname{lexmin}(s,\mathrm{payload})),\\
  \frontier_d &= \operatorname{Top}_{\beam}
  \left(U_d;\,s_d^*,\tau_L\right),
  \label{eq:beam}
\end{align}
Here $\cand_d$ contains all children, $U_d$ keeps the lowest-score candidate
for each Hash128 key $z$, and $\frontier_d$ contains the best $B$ distinct key
records.  Specifically,
$s_d^*(z)=\min\{s(x,a):z(A_a(x))=z\}$, and $\tau_L$ is the
deterministic tie order induced by a fully specified physical layout $L$,
including rank, physical buffer, and local slot.  Thus
de-duplication keeps the best candidate for each key before the global cut.
At a saturated depth, the scale follows from one multiplication:
\begin{equation}
\begin{aligned}
N_{\rm child}&=mB_{\rm eff}\\
&=24\times2{,}900{,}361{,}216\\
&=69{,}608{,}669{,}184
\quad\text{nominal logical pairs}.
\end{aligned}
\label{eq:capacity-arithmetic}
\end{equation}
The Cube4 representation uses 96 bytes per full child.  Hypothetically
materializing every child, before metadata, would require
\begin{equation}
\begin{aligned}
M_{\rm hypothetical}&=96N_{\rm child}\\
&=96\times69{,}608{,}669{,}184\\
&=6{,}682{,}432{,}241{,}664\ \text{bytes}\\
&\approx6.68\ \text{decimal TB}.
\end{aligned}
\label{eq:raw-memory}
\end{equation}

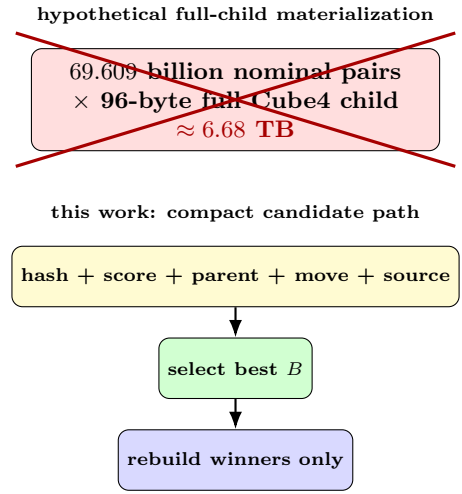
\begin{figure}[!htbp]
\centering
\begin{tikzpicture}[
  >=Latex,
  big/.style={draw,rounded corners,minimum height=1.35cm,align=center,font=\small\bfseries},
  small/.style={draw,rounded corners,minimum height=8mm,align=center,font=\scriptsize\bfseries}
]
\node[big,fill=red!13,minimum width=5.4cm] (raw)
  {$69.609$ billion nominal pairs\\$\times$ 96-byte full Cube4 child\\
   \textcolor{red!65!black}{$\approx6.68$ TB}};
\node[font=\scriptsize\bfseries,above=2mm of raw]
  {hypothetical full-child materialization};
\node[small,fill=yellow!22,below=1.25cm of raw] (record)
  {hash + score + parent + move + source};
\node[small,fill=green!18,below=4mm of record] (select) {select best $B$};
\node[small,fill=blue!15,below=4mm of select] (rebuild) {rebuild winners only};
\draw[->,thick] (record)--(select);
\draw[->,thick] (select)--(rebuild);
\node[font=\scriptsize\bfseries,above=2mm of record] {this work: compact candidate path};
\draw[very thick,red!65!black] ($(raw.north west)+(-2mm,2mm)$)--($(raw.south east)+(2mm,-2mm)$);
\draw[very thick,red!65!black] ($(raw.south west)+(-2mm,-2mm)$)--($(raw.north east)+(2mm,2mm)$);
\end{tikzpicture}
\caption{Why the system stores candidate metadata rather than every generated
state. The 6.68-TB value is hypothetical full-child materialization before
metadata, not measured device allocation. Full children are reconstructed only
after the global cut.}
\label{fig:compact-candidates}
\end{figure}

Figure~\ref{fig:compact-candidates} contrasts that hypothetical materialization
with the compact metadata path used by the system.
The system therefore never gathers the full candidate layer for a CPU transfer
or a single-GPU sort.

During a depth, candidate and frontier arrays remain in GPU memory.  The CPU
receives only $O(p)$ control counts and compact ancestry metadata.  Every GPU
therefore contributes a shard to the same next frontier; the system is not an
ensemble of independent searches.

\paragraph{Contributions.}
We distribute one logical beam without staging its frontier in CPU memory.
We state the implementation obligations needed to preserve monolithic
reduced-key top-$B$ semantics for a fixed physical layout and identity model, without
shard-local semantic caps.  We report separately an eight-H200 completed
Cube4 depth, an older eight-A100 capacity-only record, and two-T4 Megaminx
throughput.  We also report matched 1/2/4/8-RTX-3060 scaling under common and
selected execution profiles.  The implementation supports at most 128 ranks;
execution beyond eight GPUs remains unmeasured.

\begin{table*}[t]
\centering
\small
\caption{Notation legend. Read this table as the map for the equations below.}
\label{tab:notation}
\setlength{\tabcolsep}{5pt}
\begin{tabular}{@{}clclcl@{}}
\toprule
Symbol & Plain meaning & Symbol & Plain meaning & Symbol & Plain meaning \\
\midrule
$\frontier_d$ & states kept at depth $d$ & $\cand_d$ & all generated children
& $U_d$ & one best record per key \\
$B$ & desired frontier size & $m$ & moves per parent
& $p$ & number of GPUs \\
$z(x)$ & Hash128 key of state $x$ & $s(x,a)$ & integer score; smaller is better
& $T$ & current worst allowed score \\
$A_a(x)$ & child after move $a$ & $o(x)$ & GPU owning state $x$
& $\tau_L$ & physical-layout tie order \\
\bottomrule
\end{tabular}
\end{table*}

\section{Related Work and Gap}
The words \emph{large search} can refer to different quantities:
\begin{equation}
\boxed{B_{\rm live}\neq N_{\rm raw}\neq N_{\rm cumulative}
\neq N_{\rm queries}.}
\label{eq:scale-taxonomy}
\end{equation}
Here they mean retained frontier, raw candidates in one layer, states visited
over time, and independent searches.  We compare like with like.

Parallel beam search predates GPUs~\cite{anantharaman1986accelerator}.
Distributed/global CPU beams include Wijs--Lisser, a 20-million shared-memory
beam, independent MPI beams, and global or local CPU partitions
~\cite{wijs2007distributed,frohner2022parallel,kuroiwa2024parallel}.

Single-GPU systems include a speech beam, GPU A*, Cayley-graph beams,
AlphaCube, and DeepCubeA
~\cite{chong2008lvcsr,zhou2015gpuastar,soltani2026gpuastar,
chervov2025large,soibelman2025cayleypy,takano2023alphacube,
agostinelli2019deepcubea}.  Hash-distributed CPU best-first
search establishes another piece~\cite{kishimoto2013hda}.
DA* distributes A* over multiple GPUs, but does not maintain the
layer-synchronous global top-$B$ frontier studied here~\cite{he2021multigpuastar}.

QiankunNet-cuSCI combines distributed GPU de-duplication, neural inference,
and hierarchical global Top-$K$ selection and is evaluated through 64 GPUs
~\cite{sun2026cunnqssci}. Its SCI iteration updates a selected configuration
space and stages intermediate configurations through host memory; it does
not establish the all-GPU, fixed-layout layer-replacement beam contract
studied here.

Other billion-scale numbers count different objects: GPUexplore counts
cumulative reachable states~\cite{wijs2024gpuexplore}, PathWeaver serves
independent graph queries~\cite{kim2025pathweaver}, and multi-GPU decoders
usually distribute utterances, sentences, or model execution rather than one
state frontier~\cite{junczysdowmunt2016marian,braun2020gpuasr,chan2025trie}.

\begin{table*}[t]
\centering
\footnotesize
\caption{Closest prior systems under a common scale taxonomy.  ``Other scale''
is deliberately not interpreted as retained beam width.}
\label{tab:prior_art}
\setlength{\tabcolsep}{2.5pt}
\begin{tabular}{@{}p{2.7cm}p{3.7cm}p{3.0cm}p{3.0cm}p{4.2cm}@{}}
\toprule
System & Platform and search object & Retained live beam & Other scale & Missing from the claimed conjunction \\
\midrule
Wijs--Lisser 2007 & distributed CPU; one global beam & 50,000 & 135.96M cumulative states & GPU data plane and device-resident scorer \\
Chong et al. 2008 & one 8800 GTX; speech active set & not reported as retained beam & 996,903-state large model & multi-GPU state frontier \\
Frohner et al. 2022 & 46 CPU threads; one shared-memory beam & 20,000,000 & MPI: independent beams & GPU and distributed one-beam execution \\
DA* 2021 & four V100; multi-GPU A* & not a beam & up to $3\times$ versus single-GPU A* & no layer-synchronous global top-$B$ frontier \\
CayleyPy 2025 & one GPU; one Cayley-graph beam & $2^{24}$ & independent agents & multi-GPU global frontier \\
GPUexplore 3.0 & four V100; exhaustive reachable set & not a beam & $>3.15$B states & ranked score and top-$B$ \\
PathWeaver 2025 & four A6000; independent ANNS queries & not reported & 50M-vector dataset & exact layered semantics and large retained beam \\
QiankunNet-cuSCI & up to 64 GPUs; selected SCI configurations & not reported as a beam & large distributed selection workload & SCI-space update and out-of-core host staging; fixed-layout layer-replacement beam equivalence not established \\
This work & eight H200; one sharded global beam & 2,900,361,216 effective records & 69.609B nominal raw pairs/depth & H200 record: completed depth 8; full solve, implementation exactness, and H200 fixed-workload scaling unestablished; separate RTX 3060 scaling below \\
\bottomrule
\end{tabular}
\end{table*}

Table~\ref{tab:prior_art} shows the conjunction examined by the dated audit:
one physically sharded beam, GPU-resident candidate/frontier data, global key
reduction, no local semantic quota, and fixed-layout global top-$B$.  No earlier
system with the full conjunction was located in the audited set; this is not an
exhaustive proof of priority.

\FloatBarrier
\section{One Search Depth on Five GPU Streams}
One depth transforms the old frontier into the new frontier.
Figure~\ref{fig:five-stream-path} first separates the GPU-resident data path
from the compact CPU control path:
\begin{figure*}[t]
\centering
\begin{tikzpicture}[
  >=Latex,
  box/.style={draw,rounded corners,minimum width=2.25cm,minimum height=1cm,
              align=center,font=\scriptsize\bfseries},
  flow/.style={->,thick},
  meta/.style={->,dashed,gray!75}
]
\node[box,fill=blue!15] (front) {frontier\\shards};
\node[box,fill=green!18,right=4mm of front] (score) {S1: score\\S2: hash};
\node[box,fill=yellow!25,right=4mm of score] (cand) {compact\\candidates};
\node[box,fill=orange!22,right=4mm of cand] (local) {S3: local reduce\\and owner};
\node[box,fill=violet!18,right=4mm of local] (nccl) {S5: NCCL\\exchange};
\node[box,fill=red!15,right=4mm of nccl] (global) {S4: global reduce\\and threshold};
\foreach \u/\v in {front/score,score/cand,cand/local,local/nccl,nccl/global}
  \draw[flow] (\u)--(\v);
\node[box,fill=gray!12,minimum width=8cm,below=1.25cm of cand] (cpu)
  {CPU: counts and compact ancestry metadata only};
\coordinate (cpu-left) at ([xshift=9mm]cpu.north west);
\coordinate (cpu-mid) at (cpu.north);
\coordinate (cpu-right) at ([xshift=-9mm]cpu.north east);
\draw[meta] (front.south) |- (cpu-left);
\draw[meta] (local.south) -- (cpu-mid);
\draw[meta] (global.south) |- (cpu-right);
\node[font=\small\bfseries,blue!55!black,above=3mm of cand] {GPU-resident hot-data path};
\node[font=\scriptsize\bfseries,red!65!black,below=3mm of cpu]
  {No full candidate layer or full frontier is staged through CPU memory.};
\end{tikzpicture}
\caption{One search depth. Streams overlap in time; arrows show data
dependencies, not five serial phases.}
\label{fig:five-stream-path}
\end{figure*}
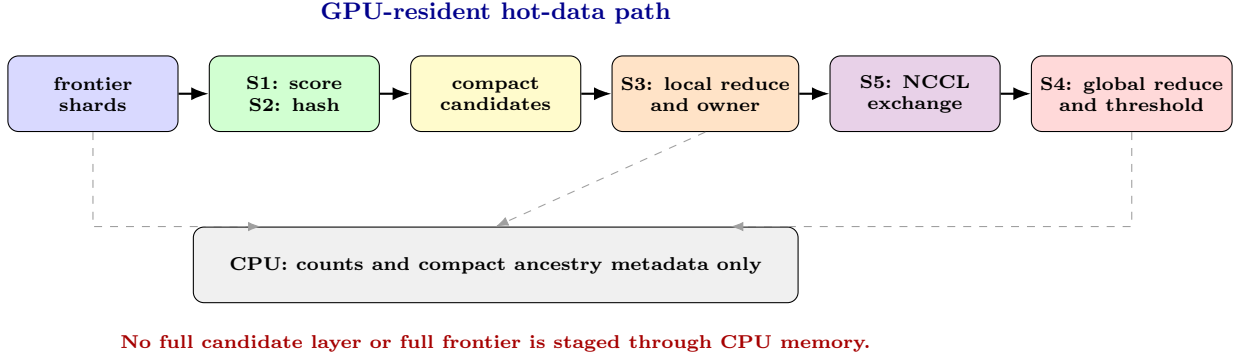

The input $\frontier_{d-1}$ is split over $p$ GPUs.  GPU $r$ owns $N_r$
parents and logically generates $mN_r$ children.  For $y=A_a(x)$, it stores a
compact candidate rather than the full child:
\begin{equation}
 c=(z(y),s(x,a),i_{\mathrm{parent}},a,r_{\mathrm{src}}),
 \label{eq:candidate}
\end{equation}
where $z(y)\in\{0,1\}^{128}$ is the child-state Hash128 key, $s(x,a)$ is a
quantized candidate score, and $r_{\mathrm{src}}$ identifies the GPU that stores
the parent.  The child state itself is reconstructed only for goal checking or
after the final selection.
We write
\begin{equation}
 x\equiv_z y \quad\Longleftrightarrow\quad z(x)=z(y).
 \label{eq:key-equivalence}
\end{equation}
All reductions and exactness statements below operate on these Hash128
key-equivalence classes.  They also describe full-state equality only if no
distinct relevant states collide under $z$.

\begin{table}[t]
\centering
\scriptsize
\caption{Compact candidate fields. Full state bytes are absent.}
\label{tab:candidate-fields}
\begin{tabular}{@{}lll@{}}
\toprule
Field & Meaning & Used for \\
\midrule
$z(y)$ & Hash128 key & key reduction, owner \\
$s(x,a)$ & integer score & ranking, threshold \\
$i_{\rm parent}$ & parent index & reconstruction \\
$a$ & applied move & reconstruction \\
$r_{\rm src}$ & source GPU & fetch and ancestry \\
\bottomrule
\end{tabular}
\end{table}

\begin{table*}[t]
\centering
\small
\caption{The five streams at a glance. ``Resident'' means that the large data
remain in GPU memory.}
\label{tab:streams}
\setlength{\tabcolsep}{5pt}
\begin{tabular}{@{}cllll@{}}
\toprule
Stream & Reads & Main operation & Writes & Why it exists \\
\midrule
1 & parent microbatch & neural or heuristic scoring & integer scores & rank candidates \\
2 & same parents & apply moves, goal test, hash & 128-bit keys & assign key classes \\
3 & scores and keys & threshold, compact, local reduce & owner buckets & cut work before exchange \\
5 & remote owner buckets & NCCL GPU-to-GPU exchange & received candidates & unite equal-key records \\
4 & owned candidates & persistent reduce and histogram & survivor shards & maintain the global cut \\
\bottomrule
\end{tabular}
\end{table*}

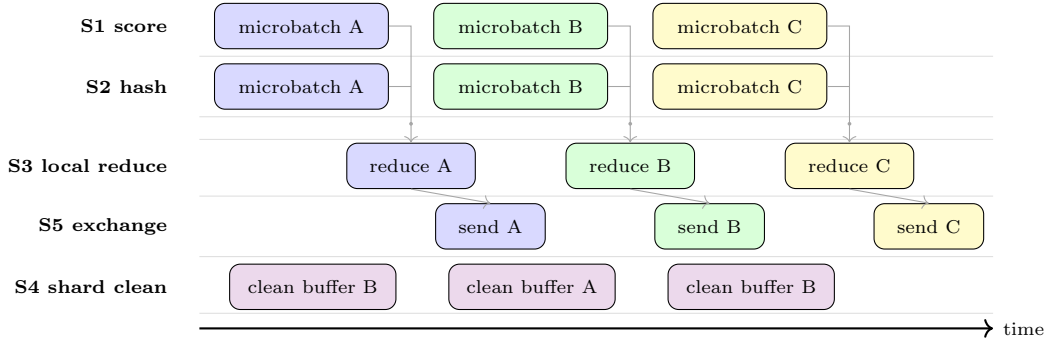
\begin{figure*}[t]
\centering
\begin{tikzpicture}[
  lane/.style={font=\scriptsize\bfseries,anchor=east},
  job/.style={draw,rounded corners,minimum height=6mm,font=\scriptsize,
             align=center},
  dep/.style={->,thin,gray!70}
]
\node[lane] at (-0.3,2.4) {S1 score};
\node[lane] at (-0.3,1.6) {S2 hash};
\node[lane] at (-0.3,0.55) {S3 local reduce};
\node[lane] at (-0.3,-0.25) {S5 exchange};
\node[lane] at (-0.3,-1.05) {S4 shard clean};
\foreach \y in {-1.4,-0.65,0.15,0.9,1.2,2.0,2.8}
  \draw[gray!25] (0,\y)--(10.5,\y);
\foreach \x/\name/\color in {0.2/A/blue!15,3.1/B/green!15,6.0/C/yellow!25}{
  \node[job,fill=\color,minimum width=2.3cm] (s1\name) at (\x+1.15,2.4) {microbatch \name};
  \node[job,fill=\color,minimum width=2.3cm] (s2\name) at (\x+1.15,1.6) {microbatch \name};
  \node[job,fill=\color,minimum width=1.7cm] (s3\name) at (\x+2.6,0.55) {reduce \name};
  \node[job,fill=\color,minimum width=1.45cm] (s5\name) at (\x+3.65,-0.25) {send \name};
}
\node[job,fill=violet!15,minimum width=2.2cm] at (1.5,-1.05) {clean buffer B};
\node[job,fill=violet!15,minimum width=2.2cm] at (4.4,-1.05) {clean buffer A};
\node[job,fill=violet!15,minimum width=2.2cm] at (7.3,-1.05) {clean buffer B};
\foreach \name in {A,B,C}{
  \coordinate (ready\name) at ($(s3\name.north)+(0,2.5mm)$);
  \draw[thin,gray!70] (s1\name.east)--(s1\name.east-|ready\name)--(ready\name);
  \draw[thin,gray!70] (s2\name.east)--(s2\name.east-|ready\name);
  \fill[gray!70] (ready\name) circle (0.7pt);
  \draw[dep] (ready\name)--(s3\name.north);
  \draw[dep] (s3\name.south)--(s5\name.north);
}
\draw[->,thick] (0,-1.6)--(10.5,-1.6) node[right,font=\scriptsize]{time};
\end{tikzpicture}
\caption{Readiness-driven overlap. Streams 1--2 process the same ring slot in
parallel; Streams 3 and 5 handle completed microbatches while Stream 4 cleans
the alternate shard buffer. The drawing shows dependencies, not measured
durations.}
\label{fig:stream-timeline}
\end{figure*}

Figure~\ref{fig:stream-timeline} then shows how readiness, rather than a fixed
serial phase order, permits overlap among the five streams.

\paragraph{Streams 1--2: score and identity.}
Parents are divided into microbatches and placed in reusable ring slots.
Streams~1 and~2 consume the same slot concurrently.  Stream~1 evaluates the
chosen scoring oracle and writes only
\begin{equation}
 s(x,a)=\operatorname{rint}_{\mathrm{RN-even}}
 \left(S\,\operatorname{clip}(\sigma(x,a),0,q_{\max})\right)
 \in\mathbb{N},
 \label{eq:score}
\end{equation}
not the floating-point output.  Here RN-even is the GPU path's default
round-to-nearest, ties-to-even mode.  The scalar path evaluates child states, whereas
the output-$m$ path obtains all $m$ action scores from each parent.  Stream~2
applies every move in registers, checks the goal, and writes the 128-bit Zobrist
hash.  Thus score evaluation and state manipulation overlap without storing the
$mN_r$ child states.

\paragraph{Stream 3: early reduction and ownership.}
When a group of ring slots has both scores and hashes, Stream~3 removes scores
above the current threshold $T$, compacts the survivors, sorts them by $z$, and
reduces equal hashes.  For each hash it retains the lexicographically smallest
pair $(s,\mathrm{payload})$, where the payload recovers the original parent and
move.  It then assigns each locally reduced candidate to
\begin{equation}
 o(x)=\operatorname{mix}(z(x))\bmod p.
 \label{eq:owner}
\end{equation}
Local candidates enter the local collector; the rest are grouped by owner.
Performing this first reduction before communication lowers traffic, but does
not replace the later global deduplication.

\paragraph{Stream 5: GPU-to-GPU exchange.}
Stream~5 exchanges the owner-grouped metadata directly between GPUs using
NCCL.  The owner rule implies
\begin{equation}
 z(x)=z(y)\Longrightarrow o(x)=o(y),
 \label{eq:owner_invariant}
\end{equation}
so, under complete and race-free routing, equal-key candidates produced by
different source GPUs meet at one owner.
Stream~5 performs communication only; received candidates re-enter the same
collector as local candidates.
The owner sets form a disjoint partition of the global reduced key set:
\begin{equation}
 U^{(r)}=\{z\in U:o(z)=r\},\qquad
 U=\biguplus_{r=0}^{p-1}U^{(r)}.
 \label{eq:owner-partition}
\end{equation}

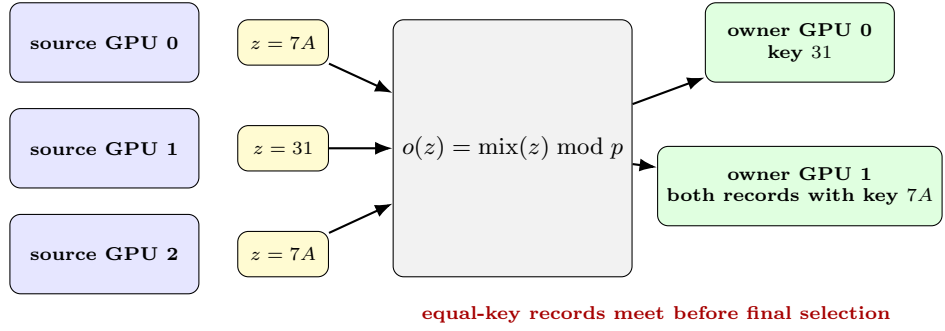
\begin{figure*}[t]
\centering
\begin{tikzpicture}[
  >=Latex,
  gpu/.style={draw,rounded corners,minimum width=2.5cm,minimum height=1.05cm,
              align=center,font=\scriptsize\bfseries},
  key/.style={draw,rounded corners,fill=yellow!22,font=\scriptsize,
              minimum width=1.2cm,minimum height=6mm},
  route/.style={->,thick}
]
\node[gpu,fill=blue!10] (s0) at (0,1.4) {source GPU 0};
\node[gpu,fill=blue!10] (s1) at (0,0) {source GPU 1};
\node[gpu,fill=blue!10] (s2) at (0,-1.4) {source GPU 2};
\node[key,right=5mm of s0] (a0) {$z=7A$};
\node[key,right=5mm of s1] (b1) {$z=31$};
\node[key,right=5mm of s2] (a2) {$z=7A$};
\node[draw,rounded corners,fill=gray!10,minimum width=2.5cm,minimum height=3.4cm,
      align=center,font=\small\bfseries] (owner) at (5.4,0) {$o(z)=\operatorname{mix}(z)\bmod p$};
\node[gpu,fill=green!12] (d0) at (9.2,1.4) {owner GPU 0\\key $31$};
\node[gpu,fill=green!12] (d1) at (9.2,-0.5) {owner GPU 1\\both records with key $7A$};
\draw[route] (a0)--(owner);
\draw[route] (b1)--(owner);
\draw[route] (a2)--(owner);
\draw[route] (owner)--(d0);
\draw[route] (owner)--(d1);
\node[font=\scriptsize\bfseries,red!65!black,align=center]
  at (7.3,-2.2)
  {equal-key records meet before final selection};
\end{tikzpicture}
\caption{Required owner routing in the abstract algorithm. Equal 128-bit keys
choose the same destination; a conforming owner reduction then keeps one
representative per key. Local reduction cuts traffic. This establishes neither
unconditional full-state equality nor compliance of the audited CUDA revision.}
\label{fig:owner-routing}
\end{figure*}

Figure~\ref{fig:owner-routing} illustrates the required meeting point for
equal-key records produced on different source GPUs.

\paragraph{Stream 4: persistent survivor set.}
The collector partitions owned candidates into logical shards.  Each shard has
two resident buffers: Stream~3 appends to one while Stream~4 cleans the other.
A Stream~4 job merges the clean and newly appended regions, applies its snapshot
of $T$, compacts, sorts by $z$, and retains the best metadata for each hash.
There is no per-shard top-$k$ and no semantic shard quota: a shard may retain
every candidate below $T$.  Consequently sharding does not approximate the
score cut; only the equal-score boundary order is defined by the fixed layout.

\subsection{Work, Memory, and Frontier Scaling}
Let $a$ denote the branching factor (the earlier move count $m$), $G$ the
number of ranks, $B_d=|\frontier_d|\leq B$, and
\begin{equation}
 N_d=aB_{d-1};\qquad N=aB\quad\text{at saturation}.
 \label{eq:complexity-candidates}
\end{equation}
Let $\ell$ be the logical state length, $s_{\rm state}$ the bytes per stored
state, and $m_{\rm meta}$ the bytes per candidate metadata record.  These names
avoid conflict with the score-quantization scale $S$ in
Eq.~\eqref{eq:score}.  Let $f_{\rm child}$ be the cost of one scalar child
score and $f_{\rm parent}(a)$ the cost of one parent inference returning $a$
action scores. Aggregate base move-generation and hashing work is
$\Theta(N_d\ell)$. This assumes fixed bounded goal-check/suffix work;
a varying suffix neighborhood of size $R_{\rm suffix}$ can add
$O(N_dR_{\rm suffix}\ell)$ work. Scoring work is
\begin{equation}
 W_{\rm score}=\begin{cases}
 \Theta(N_df_{\rm child}),&\text{scalar child scorer},\\
 \Theta(B_{d-1}f_{\rm parent}(a)+N_d),&\text{output-$a$ scorer}.
 \end{cases}
 \label{eq:scorer-complexity}
\end{equation}
The second line includes writing and quantizing the $a$ outputs; it is not $a$
independent neural-network calls per parent.

For orientation, a \emph{sort-based abstract reference} that comparison-sorts
all implicit candidates has
\begin{align}
 W_{\rm ref}&=O(N_d\ell+W_{\rm score}+N_d\log N_d),\nonumber\\
 M_{\rm ref}&=O(B_{d-1}s_{\rm state}+N_dm_{\rm meta}).
 \label{eq:reference-complexity}
\end{align}
This is one exact-beam implementation, not an intrinsic lower bound and not a
bound claimed for a historical dispatcher.  Hash tables, deterministic
selection, or fixed-score histograms have different work bounds.  Streaming
avoids $\Theta(N_ds_{\rm state})$ full-child storage, but all $N_d$ metadata
records may survive the threshold; therefore it does not imply worst-case
$O(B)$ metadata independent of $a$.

\begin{figure}[!ht]
\centering
\begin{tikzpicture}[
  card/.style={draw,rounded corners,minimum width=3.15cm,minimum height=1.55cm,
               text width=2.85cm,align=center,font=\scriptsize,inner sep=2mm},
  title/.style={font=\small\bfseries}
]
\node[title] at (1.775,2.05) {What can stop more GPUs from helping?};
\node[card,fill=blue!10] (scorelimit) at (0,0.85)
  {\textbf{Scoring work}\\scorer throughput\\diagnostic: scorer time};
\node[card,fill=orange!14] (sortlimit) at (3.55,0.85)
  {\textbf{Sort and reduction}\\capacity sorts; repeated jobs\\diagnostic: $|\mathcal J_4|,\ C_k$};
\node[card,fill=green!12] (netlimit) at (0,-1.15)
  {\textbf{Owner exchange}\\traffic and skew\\diagnostic: \mbox{remote bytes};\\$\max_r|U^{(r)}|$};
\node[card,fill=violet!12] (memlimit) at (3.55,-1.15)
  {\textbf{Live memory}\\phase peak; history/drain\\diagnostic: peak bytes;\\$t_{\rm service},t_{\rm final}$};
\end{tikzpicture}
\caption{Conceptual diagnostic map, not a measured timeline or an established
GPU-count crossover.  The boxes are independent checks: overlap requires
resource independence, per-rank division requires balance, and a bounded-pass
count for Stream~4 jobs is not established by the present audit.}
\label{fig:scaling-limits}
\end{figure}
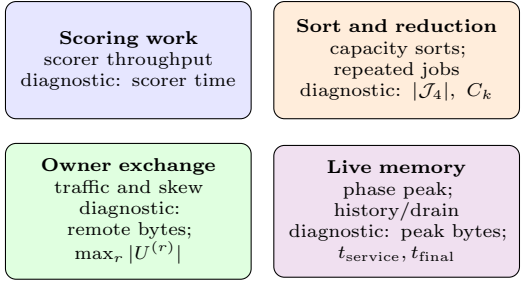

Figure~\ref{fig:scaling-limits} is a diagnostic checklist for interpreting
these counts.  Its boxes are possible limiting resources, not measured shares
of runtime or a predicted crossover GPU count.

The read-only audit of current dirty HEAD
\texttt{81ed9f0287c1f9057648119a253f208c31f7a872} finds full-capacity CUB
comparison sorts rather than a linear Hash128 radix reduction.  With $Q$ the
Stream~3 chunk capacity, $q_r$ useful chunks on rank $r$, and
$U_3\leq N_d$ locally reduced survivors across chunks, its accounted work is
\begin{equation}
 W_3=O\!\left(\sum_r q_rQ\log Q+N_d+GU_3\right),
 \label{eq:stream3-complexity}
\end{equation}
plus padded and service rounds.  The $GU_3$ term is from per-peer owner scans.
For Stream~4 jobs $k$ with sorted capacity $C_k$ and $H$ score bins,
\begin{equation}
 W_4=O\!\left(\sum_{k\in\mathcal J_4}C_k\log C_k
              +\sum_{k\in\mathcal J_4}(C_k+H)\right).
 \label{eq:stream4-complexity}
\end{equation}
No bounded-pass or amortized charging argument for the job count is established;
hence Eq.~\eqref{eq:stream4-complexity} is not simplified to $O(N_d)$ or
$O(N_d\log N_d)$.  These current-code formulas do not transfer correctness or
complexity to historical pin \texttt{a1db0e6d9bb5458c8a842b37dfa99572d3025667}
or to the H200, two-T4, or RTX 3060 measured working snapshots without
per-file comparison.  Appendix~\ref{app:complexity} records the remaining
implementation-specific terms.

\begin{table}[!ht]
\centering
\footnotesize
\caption{Scaling with frontier width.  The right column requires balanced
parents, owners, admitted work, and adequate capacity; it is not guaranteed.}
\label{tab:complexity-scaling}
\setlength{\tabcolsep}{3pt}
\begin{tabular}{@{}p{1.55cm}p{2.35cm}p{2.55cm}@{}}
\toprule
Quantity & Aggregate or worst case & Balanced conditional per rank \\
\midrule
Candidates & $N_d=aB_{d-1}$ & $N_d/G$ \\
Frontier states & $\Theta(B_ds_{\rm state})$ & $\Theta(B_ds_{\rm state}/G)$ \\
Survivor metadata & $O(N_dm_{\rm meta})$; one owner may receive $\Theta(N_d)$ & $O(N_dm_{\rm meta}/G)$ \\
Candidate exchange & $O(N_dm_{\rm meta})$ bytes & about aggregate$/G$, plus latency \\
Final exchange & $O(B_d(m_{\rm meta}+s_{\rm state}))$ bytes & balanced only after redistribution \\
History through $D$ & $O(\!\sum_d B_dh)$; saturated $O(DBh)$ & host/disk service, not GPU balance \\
\bottomrule
\end{tabular}
\end{table}

Peak GPU memory also includes allocated frontier capacity, rings, both
accumulator buffers, comparison-sort temporary storage, model/scorer workspace,
and communication/runtime workspace.  Mutually exclusive allocator phases are
combined by a maximum, not by summing every phase.  Across depths the work is
$\sum_{d=1}^{D}W_d$, not $a^D$ once the beam is saturated.  Finally, five CUDA
streams imply neither a fivefold speedup nor linear strong scaling.  Under
verified balance, bandwidth, and resource independence one may model a steady
pipeline as
\begin{equation}
 T_d\approx t_{\rm fill}+\max_i t_i+t_{\rm drain/final},
 \label{eq:pipeline-time-model}
\end{equation}
but skew, peer latency, collectives, HBM contention, history I/O, and the final
drain are non-scaling floors.  Equation~\eqref{eq:pipeline-time-model} is a
conditional performance model, not a theorem or a new benchmark result.

\paragraph{Classical top-$\beam$ versus monotone thresholding.}
Let $C$ be the complete candidate multiset, write $s(c)$ for the quantized
integer score in Eq.~\eqref{eq:score}, and let
$U=\operatorname{Reduce}_{z}(C;\operatorname{lexmin}(s,\mathrm{payload}))$ be
its one-entry-per-key reduction, carrying
$s^*(z)=\min_{c:z(c)=z}s(c)$ and the associated best payload.  For fixed layout
$L$, the complete ordering key is
\begin{equation}
 \kappa_L(z)=\bigl(s^*(z),\tau_L(z)\bigr),
 \label{eq:ordering-key}
\end{equation}
where smaller score is better and $\tau_L$ is the deterministic prefix order
over rank, physical shard/buffer, and local slot for equal scores.  Thus $L$
specifies the realized physical arrangement, not merely GPU and shard counts.
A classical reference
materializes $U$, sorts it by $\kappa_L$, and returns, for
$b=\min(\beam,|U|)$,
\begin{equation}
 \operatorname{Top}_{\beam}(U)=
 \begin{cases}
 \varnothing,&b=0,\\
 \{z\in U:\kappa_L(z)\leq \kappa_{L,(b)}(U)\},&b>0.
 \end{cases}
 \label{eq:classical_topb}
\end{equation}
Thus the order statistic is used only when $b>0$.  The streamed abstract
algorithm need not materialize $U$.  At update $j$, let $W_j$ be the entries
counted by the threshold snapshot.  The safety premise is witness-based: each
entry $(z,q)\in W_j$ represents a different key and has an observed integer
score $q\geq s^*(z)$.  Entries may be omitted, and a stored score may be stale
and higher than its eventual minimum; either case can only loosen the cut.
With $H_j$ the histogram of witness scores, define
\begin{align}
 \widehat T_j&=\min\left\{t:\sum_{u=0}^{t}H_j(u)\geq\beam\right\},\nonumber\\
 T_j&=\min\{T_{j-1},\widehat T_j\},\qquad T_0=+\infty.
 \label{eq:online_threshold}
\end{align}
When $|W_j|<\beam$, $\widehat T_j=+\infty$.
Streams~3 and~4 may discard only candidates with $s(c)>T_j$; the score boundary
$s(c)=T_j$ is retained because its final membership depends on $\tau_L$.

\begin{figure*}[t]
\centering
\begin{tikzpicture}[
  cell/.style={draw,minimum width=9mm,minimum height=6mm,font=\scriptsize},
  row/.style={font=\small\bfseries,anchor=east},
  note/.style={font=\scriptsize}
]
\node[font=\small\bfseries] at (4,2.55) {fixed beam size $B=7$ in every row};
\node[row] at (-0.85,1.8) {early};
\node[row] at (-0.85,0.8) {middle};
\node[row] at (-0.85,-0.2) {final};
\foreach \x/\v in {0/1,1/2,2/3,3/4,4/5,5/6}
  \node[cell,fill=green!18] at (\x,1.8) {\v};
\node[cell,fill=yellow!30] at (6,1.8) {7};
\node[cell,fill=yellow!30] at (7,1.8) {7};
\node[cell,fill=gray!18] at (8,1.8) {9};
\foreach \x/\v in {0/0,1/1,2/2,3/3,4/4,5/5}
  \node[cell,fill=green!18] at (\x,0.8) {\v};
\node[cell,fill=yellow!30] at (6,0.8) {6};
\node[cell,fill=yellow!30] at (7,0.8) {6};
\node[cell,fill=gray!18] at (8,0.8) {8};
\foreach \x/\v in {0/0,1/1,2/2,3/3,4/4}
  \node[cell,fill=green!18] at (\x,-0.2) {\v};
\node[cell,fill=yellow!30] at (5,-0.2) {5};
\node[cell,fill=yellow!30] at (6,-0.2) {5};
\node[cell,fill=yellow!30] at (7,-0.2) {5};
\node[cell,fill=gray!18] at (8,-0.2) {7};
\node[note,red!65!black] at (9.35,1.8) {$T_1=7$};
\node[note,red!65!black] at (9.35,0.8) {$T_2=6$};
\node[note,red!65!black] at (9.35,-0.2) {$T^*=5$};
\draw[->,very thick,red!65!black] (10.2,1.55)--node[right,note]{never rises}(10.2,0.05);
\node[note,green!45!black,align=center] at (1,-1.0) {keep for now};
\node[note,orange!70!black,align=center] at (4.8,-1.0)
  {retain boundary until the\\physical-layout tie break};
\node[note,gray!70!black,align=center] at (8.1,-1.0) {safe to discard};
\node[note] at (4.9,-1.75)
  {Final row: $n_{<}=5$; $n_{=}=2$ of three score-5 keys win by $\tau_L$.};
\end{tikzpicture}
\caption{Monotone thresholding for one fixed $B=7$ with integer scores.
Green entries are retained provisionally, not guaranteed final membership.
Values above the current threshold cannot return; every equal-score boundary
entry remains until the fully specified physical-layout tie-break. Numbers are
illustrative.}
\label{fig:monotone-threshold}
\end{figure*}
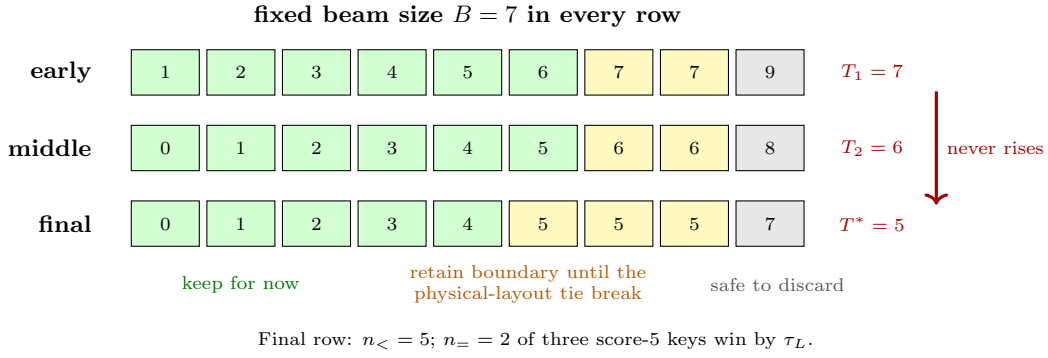

Figure~\ref{fig:monotone-threshold} visualizes the conservative integer-score
cut and the retained equal-score boundary.

\paragraph{Lemma 1 (the published threshold never rises).}
The update rule publishes a minimum,
\begin{equation}
T_j=\min\{T_{j-1},\widehat T_j\}\leq T_{j-1}.
\label{eq:threshold-monotone}
\end{equation}
This statement does not require a physical snapshot to equal
$\operatorname{Reduce}_z(C_j)$ (Fig.~\ref{fig:monotone-threshold}).

\paragraph{Lemma 2 (a discarded candidate cannot win the reduced-key beam).}
If $s(c)>T_j$, some published witness snapshot has at least $\beam$ distinct
keys with observed scores at most $T_j$.  Their eventual minimum scores are no
larger than those observations, hence are strictly better than $s(c)$.  Thus
$c$ cannot represent a key in the final top-$\beam$.  Omitted witnesses or
stale higher scores make $T_j$ weaker, not unsafe.  Counting one key twice,
however, destroys the distinct-key premise.

\paragraph{Theorem (conditional abstract reduced-key equivalence).}
For the same complete candidate set, integer quantization, Hash128 mapping
$z$, payload order, and fully specified physical layout $L$, Lemmas 1--2 give the same selected
key representatives if every candidate is processed, all records for one key
are globally reduced before contributing distinct threshold witnesses,
rejection is strict above the committed threshold, no local capacity silently
drops an eligible record, and exchange plus final physical-prefix boundary
selection complete:
\begin{equation}
 \frontier_{d,z}^{\rm distributed}(L)
 =\frontier_{d,z}^{\rm monolithic}(L).
 \label{eq:exactness-theorem}
\end{equation}
The equality concerns selected membership under the same $L$.  It does not
assert an identical serialized output order after redistribution.  Fixed GPU
and shard counts alone are insufficient if scheduling changes physical-buffer
or local-slot placement.  This is a theorem about the abstract algorithm, not
by itself a proof that a particular CUDA revision satisfies every obligation.

\paragraph{Pinned-code correspondence.}
The historical one-T4 measurement is pinned to
\texttt{a1db0e6d9bb5458c8a842b37dfa99572d3025667}.  Its static audit
verifies Hash128 owner calculation, per-buffer lexicographic reduction,
inclusive threshold retention, monotone threshold publication, and final
score-boundary phases.  It does not establish the full theorem correspondence.
Stream~4 reduces each physical sibling buffer separately, while
\texttt{threshold.cu:34--61} combines per-physical-shard snapshots and the
final-selection paths scan all physical buffers; no global
cross-buffer key reduction was found.  In addition, the multi-rank Stream~3
owner scatter reads and compacts the same metadata array in one kernel across
peer blocks without a grid-wide synchronization point.  These are static
audit findings, not reproduced GPU failures.  The scatter issue is a
\texttt{world\_size}$>1$ code-path risk and is not evidence that the one-T4
measurement failed.  Cross-buffer duplication would violate the distinct-key
witness premise.  Final ties use rank/physical-buffer/local-slot prefixes
(\texttt{threshold.cu:147--154,214--229,252--279} and
\texttt{dispatcher.cu:4480--4494}).  The H200 revision and historical two-T4 \texttt{v118} run
inherit neither these defects nor correctness without their own code
comparison.

\paragraph{Identity and exactness scope.}
The word \emph{exact} means beam-selection exactness and has three explicit
boundaries:
\begin{equation}
\boxed{\begin{gathered}
\text{quantized integer scores}\\
+\ \text{Hash128 key equivalence}\\
+\ \text{fully specified physical-layout ties}
\end{gathered}}
\label{eq:exactness-scope}
\end{equation}
Compact candidates represent the equivalence classes of
Eq.~\eqref{eq:key-equivalence}; they do not carry the full state representation.  If no
distinct relevant full states share a Hash128 key, the key-level equality also
implies full-state selection equality.  Otherwise, a collision merges those
states within the stated model.  Boundary ties use the prefix over rank,
physical buffer, and local slot.  Changing that arrangement may change selected
membership; fixed GPU/shard counts do not imply schedule-independent results.
The theorem does not claim a canonical output order.  It concerns beam
selection, not shortest-path optimality.  Historical measurements remain tied
to their own revisions and do not inherit correctness or defects from this
theorem or from another source tree.

\paragraph{Final cut and materialization.}
After all five streams are drained, the ranks combine the final histograms.  If
$|U|\geq\beam$, they choose the smallest $T^*$ such that
\begin{equation}
 \sum_{u=0}^{T^*}H(u)\geq\beam.
 \label{eq:threshold}
\end{equation}
All states below $T^*$ survive, and the same physical-prefix tie rule cuts its
boundary to exactly $\beam$ selected members.  Define
\begin{equation}
 n_{<}=\sum_{u<T^*}H(u),\qquad n_{=}=\beam-n_{<}.
 \label{eq:boundary-count}
\end{equation}
The final cut keeps every key below $T^*$ and the first $n_=$ boundary keys by
$\tau_L$.  This defines membership, not the later output serialization order.
If $|U|<\beam$, all available keys survive instead.  Therefore
\begin{equation}
 |\frontier_d|=\min(\beam,|U|).
 \label{eq:frontier-size}
\end{equation}
The
selected metadata are load-balanced across GPUs.
Only then does each source GPU reconstruct the requested children and send
them directly to their target GPUs.  The CPU receives $O(p)$ count vectors plus
compact selected-candidate metadata used to reconstruct ancestry, but no full
state or frontier arrays.

The five streams are readiness-driven rather than five serial phases.  Streams
1--2 fill ring slots almost continuously while Streams~3 and~4 reduce earlier
batches and Stream~5 exchanges remote survivors.  The mathematical obligation
is that every published threshold count is a distinct-key witness and every
final key class is globally reduced before the cut.  Under that premise, a
conforming implementation realizes the same selected reduced-key membership
as its monolithic reference for the same physical tie layout, without
materializing the raw candidate layer.  The pinned-code gaps above must be
closed before making that statement unconditionally for a specific measured
binary.

\FloatBarrier
\section{Experiments}
\paragraph{Protocol.}
Historical records below retain their stated single-run or summary scope.  The
new RTX 3060 scaling points are medians of three timed repeats after a full
warmup; selected profiles were re-warmed and repeated independently.  We
report computational-cost measurements separately from solution-quality
evidence. For each result we state the hardware, workload, model, and
timing boundary to the extent preserved by its own record; missing fields
are not transferred from another run or revision.  End-to-end (E2E) time
includes generation, de-duplication, communication, selection, and history;
it must not be confused with an isolated scoring-kernel rate.

Figure~\ref{fig:claim-map} separates the three legacy headline evidence records
before the detailed platform tables.
\begin{figure*}[t]
\centering
\begin{tikzpicture}[
  claim/.style={draw,rounded corners,minimum width=4.8cm,minimum height=1.65cm,
                align=center,font=\small},
  head/.style={font=\small\bfseries}
]
\node[claim,fill=blue!10] (cap) at (0,0)
 {\textbf{Completed depth}\\$8\times$H200, Cube4\\$B_{\rm eff}=2.900361216$B\\$931.266$ s};
\node[claim,fill=green!12] (thr) at (5.4,0)
 {\textbf{Throughput}\\$2\times$T4\\$30.274$ M children/s\\no scaling beyond two};
\node[claim,fill=yellow!18] (sol) at (10.8,0)
 {\textbf{Solution}\\Megaminx superflip\\51 replay-valid moves\\no optimality proof};
\node[font=\scriptsize\bfseries,red!65!black,below=4mm of thr]
 {Three claims, three evidence records; missing fields are never transferred between them.};
\end{tikzpicture}
\caption{The three legacy headline results use independent evidence. The H200 Cube4
depth, T4 Megaminx throughput, and replayed solution answer different
questions and are not a scaling series.}
\label{fig:claim-map}
\end{figure*}

\begin{table*}[t]
\centering
\footnotesize
\caption{Measured performance by platform.  E2E is the production critical
path; micro is an isolated kernel pipeline; capacity has no timing claim.
n/d: not derived; n/a: not applicable; n/r: not recorded.}
\label{tab:platforms}
\setlength{\tabcolsep}{3pt}
\begin{tabular}{@{}p{2.0cm}p{2.4cm}p{3.8cm}p{3.2cm}p{1.8cm}p{1.8cm}@{}}
\toprule
Platform & Workload & Beam $B$ & Measurement & s/depth & Raw child rate \\
\midrule
RTX 3070 8GB & puzzle 0, depth 20 & $2^{22}$ & E2E complete; $60.623$ s total & 3.031 avg. & n/d \\
RTX 3070 8GB & Stream 1, $B_\mu=4096$, $c=3$ & --- & micro complete & n/a & $48.142$ M/s \\
Kaggle $2\times$T4 & puzzle 20, radius 4, limit 72 & $B_{\rm req}=82{,}615{,}524$; $B_{\rm eff}=82{,}837{,}504$ & solved; $350.402$ s; CSV+summary & 65.670 saturated & $30.274$ M/s derived \\
$8\times$A100 40GB & puzzles 900/950/1000, depth 8 & $B_{\rm req}=770{,}883{,}178$; $B_{\rm eff}$ not recorded & stable; $39{,}745$ MiB/GPU; summary & n/r & n/r \\
$8\times$H200 & Cube4 puzzle 1000, depth 8 & $B_{\rm req}=2.900$B; $B_{\rm eff}=2{,}900{,}361{,}216$ & completed depth; $141{,}820$ MiB/GPU observed & 931.266 & $74.746$ M/s derived \\
\bottomrule
\end{tabular}
\end{table*}

\begin{table*}[t]
\centering
\small
\caption{How to read the legacy headline results. A derived value is shown only when
all inputs come from the same run.}
\label{tab:evidence-boundaries}
\begin{tabular}{@{}p{3.0cm}p{4.2cm}p{4.0cm}p{4.8cm}@{}}
\toprule
Run & Directly recorded & Exact derivation & Not established \\
\midrule
$2\times$T4 throughput & $B_{\rm req}$, $B_{\rm eff}$, depth times, wall time
& $24B_{\rm eff}/65.670=30.274$ M children/s
& throughput beyond two GPUs \\
$8\times$A100 capacity & $B_{\rm req}$, depth 8, peak MiB/GPU, stable status
& $24B_{\rm req}\geq18.501$ B children/depth
& wall time, exact $B_{\rm eff}$, speed, strong scaling \\
$8\times$H200 Cube4 & all eight rank logs complete depth 8; effective width;
device telemetry & $24B_{\rm eff}=69{,}608{,}669{,}184$ nominal pairs;
$74.746$ M/s derived & solved puzzle, depth-9 time, true maximum, CUDA-OOM boundary,
or cross-platform scaling \\
RTX 3070 E2E & 20 completed depths, $60.623$ s total
& $60.623/20=3.031$ s/depth average
& saturated child rate \\
RTX 3070 micro & isolated Stream~1 counters and time
& $48.142$ M candidates/s
& full-system throughput \\
\bottomrule
\end{tabular}
\end{table*}

Tables~\ref{tab:platforms} and~\ref{tab:evidence-boundaries} separate legacy records,
derivations, and missing measurements.  Empty cells mean ``not measured.''
The T4 and H200 rates count nominal parent--generator pairs before
de-duplication.  No RTX 3070 child rate is derived because its early depths are
not saturated.

For the saturated T4 depths, the recorded mean and derived child rate are
\begin{align}
 \bar t_{7:10}&=\frac{1}{4}\sum_{d=7}^{10}t_d=65.670\ \mathrm{s},
 \label{eq:t4-mean-time}\\
 R_{\rm child}&=\frac{mB_{\rm eff}}{\bar t_{7:10}}
 =\frac{24\times82{,}837{,}504}{65.670}
 =30.274\times10^6\ \mathrm{s}^{-1}.
 \label{eq:t4-throughput}
\end{align}

\paragraph{Local RTX 3070.}
The folded-input half2 kernel raised isolated Stream~1 throughput from about
$24.7$ to $48.142$ million candidates/s.  Over 1,202 Nsight launches, its time
fell from $9.678$ to $3.232$ s ($2.99\times$).  On the same complete depth-20
workload, E2E wall time fell from $115.334$ to $60.623$ s ($1.90\times$).  An
older run that failed at depth 33 is excluded.

\paragraph{Kaggle $2\times$T4.}
Two T4 ranks jointly maintain one beam.  The requested
$82{,}615{,}524$ states aligned to $B_{\rm eff}=82{,}837{,}504$; the run solved
in $350.402$ s with 128 shards and $B_\mu=4096$.  At $B=2^{26}$, the reported saturated-depth means for the tested 8-shard
and 64-shard configurations were $69.365$ and $51.732$ s, respectively
($1.34\times$).  The isolated $40.60$ M/s scoring rate is not an E2E rate.

\paragraph{$8\times$A100 capacity-only record.}
The run establishes capacity at requested width
$B_{\rm req}=770{,}883{,}178$.  Its summary records stable depth-8 execution
on three Megaminx instances and $39{,}745$ MiB peak memory per GPU.  It does
not record wall time, exact aligned width, or a complete runtime manifest.
The pinned \texttt{2d5f978fcfc1f018813fc49e9d9aeec32ea5e22c} launcher targets
eight 80GB A100s and a requested 1.4B width; it does not reconstruct this
missing 40GB run.  The implementation limit is 128 ranks.

\paragraph{$8\times$H200 Cube4 completed depth.}
On one Vast instance, eight H200 GPUs (143,771 MiB each) ran Cube4 puzzle 1000
with 24 generators.  Requested width $2{,}900{,}000{,}000$ aligned to
$B_{\rm eff}=2{,}900{,}361{,}216=8\times362{,}545{,}152$.  All eight rank logs
record completed depth 8; using the conservative maximum gives $931.266$ s.
Thus the saturated layer represents
\begin{equation}
 24B_{\rm eff}=69{,}608{,}669{,}184
 \quad\Longrightarrow\quad
 R_{\rm pair}=74.746\times10^6\ \mathrm{s}^{-1},
 \label{eq:h200-throughput}
\end{equation}
where the rate is derived, not a native distinct-state counter.  Observed
device-wide peak usage was 141,820 MiB/GPU.  The solver revision was
\path{3b756afe9e7636e53520fa6de7e60a167bbaa79a}.  The run was a throughput benchmark evaluated at a completed saturated depth.  The 2.94B attempt failed during NCCL
count exchange and the 3.00B attempt failed a static budget gate; neither
establishes a true maximum or CUDA-OOM boundary.  This Cube4/H200 run and the
Megaminx/T4 run are not a strong-scaling comparison.

\input{rtx3060_scaling}

\section{Comparison with CayleyPy and Megaminx Result}
Comparisons answer different questions and must not be mixed:
\begin{equation}
\boxed{\text{capacity}\neq\text{speed}\neq\text{solution quality}.}
\label{eq:comparison-types}
\end{equation}
CayleyPy shows that wider beams can improve solution quality on one GPU.  In
its $S_{20}$ experiment, width grows from $2^{10}$ (no solution) to $2^{20}$
(length 190); its large-cube agents use $2^{24}$
~\cite{soibelman2025cayleypy,chervov2025large}.  The completed H200 depth has
$B_{\rm eff}=2{,}900{,}361{,}216=172.875\times2^{24}$ effective records.  This
is only a reported width ratio: it is not a speed, solution-quality, scaling,
or verified unique-state comparison.

\begin{table*}[t]
\centering
\small
\caption{Which comparisons are valid. A speed ratio requires the same model
contract and workload; a capacity ratio does not imply a speed ratio.}
\label{tab:comparison-map}
\begin{tabular}{@{}p{3.4cm}p{4.5cm}p{4.4cm}p{4.3cm}@{}}
\toprule
Pair & Matched quantities & Valid statement & Invalid inference \\
\midrule
MultiGPU vs CayleyPy scalar & checkpoint, model head, moves, T4, widths
& recorded search-time ratio at a shared width (with stated timer boundaries) & output-24 or node-rate ratio \\
MultiGPU vs Pilgrim Cube4 & state, Q checkpoint, GPU, $B$, depth 10
& depth-10 pair rate & memory ratio across different counters \\
MultiGPU vs AlphaCube & native released systems only
& execution and failure-mode inventory & direct speedup \\
MultiGPU vs DeepCubeA & hardware-side native run
& native A* iteration and node statistics & retained-beam capacity \\
Eight H200 completed depth vs CayleyPy & effective-record and reported beam-width definitions
& $172.875\times$ width ratio & speed, quality, scaling, or verified unique-state ratio \\
\bottomrule
\end{tabular}
\end{table*}

\begin{table}[t]
\centering
\small
\caption{Published CayleyPy beam-width trend (Table 8 of
\cite{soibelman2025cayleypy}).}
\label{tab:cayleypy_width}
\begin{tabular}{@{}rrrrrr@{}}
\toprule
$W$ & $2^{10}$ & $2^{12}$ & $2^{14}$ & $2^{16}$ & $2^{20}$ \\
\midrule
Length & --- & 252 & 202 & 194 & 190 \\
\bottomrule
\end{tabular}
\end{table}

We additionally measured Pilgrim and our search implementation on the same
Cube4 state, checkpoint, single GPU, beam, and ten layers.  The
3,383,064-parameter scorer maps 96 six-class input positions to 56 piece tokens
plus CLS (57 tokens), applies input layer normalization, then four pre-LN
Transformer blocks with width 256, eight heads, and a 1024-wide feed-forward
layer, followed by a 24-output head.  Inference uses FP16 and microbatches of
384.  Table~\ref{tab:cube4_pair} reports the tenth layer, where every saturated
frontier expands $24B$ pairs.

\begin{table}[t]
\centering
\small
\setlength{\tabcolsep}{2pt}
\caption{Matched Cube4 depth-10 computation at $B=4{,}194{,}304$. Ours pair rates are derived from the recorded rounded depth times and rounded to the nearest pair/s.}
\label{tab:cube4_pair}
\resizebox{\columnwidth}{!}{%
\begin{tabular}{@{}llrrr@{}}
\toprule
GPU & Implementation & s/depth (depth 10) & Pairs/s & Peak VRAM \\
\midrule
T4 & Pilgrim & 198.153 & 508,007 & 2.90 GiB reserved \\
T4 & Ours & 181.282 & $\approx555{,}286$ & 1,949 MiB used \\
P100 & Pilgrim & 314.487 & 320,087 & 2.90 GiB reserved \\
P100 & Ours & 301.955 & $\approx333{,}372$ & 1,949 MiB used \\
\bottomrule
\end{tabular}%
}
\end{table}

\subsection{Native approach inventory and matched model families on one T4}
Table~\ref{tab:four_t4} is an inventory, not a speed ranking.  The Megaminx
tests form two separate families: scalar output-1 and Q output-24.  A ratio is
valid only when checkpoint, architecture, head, state, moves, depth, width,
dtype, and hardware match.  CayleyPy cannot consume the output-24 head.  A
node rate is reported only when the native solver exposes a valid counter; it
is never replaced by $B/t$.

\begin{table*}[t]
\centering
\small
\setlength{\tabcolsep}{3pt}
\caption{Model contracts used by the reported tests. Parameter counts refer
to the trainable source model before batch-normalization folding. Full layer
sequences remain in the artifact tables.}
\label{tab:model_architectures}
\begin{tabular}{@{}p{3.4cm}rp{2.1cm}p{6.7cm}@{}}
\toprule
Model family & Parameters & Output & Compatible implementations \\
\midrule
Cube3 AlphaCube \texttt{large} & 118,939,666 & 18 actions & AlphaCube only \\
Cube3 DeepCubeA \texttt{cube3} & 14,663,001 & scalar & DeepCubeA only \\
Megaminx scalar MLP & 34,766,849 & scalar & Pilgrim Searcher, CayleyPy, MultiGPUBeamSearch \\
Megaminx QMLP2RB & 23,978,008 & 24 actions & Pilgrim QSearcher, MultiGPUBeamSearch \\
Cube4 Q Transformer & 3,383,064 & 24 actions & matched Pilgrim/MultiGPU Cube4 and $8\times$H200 tests \\
\bottomrule
\end{tabular}
\end{table*}

Matched implementations load identical checkpoint bytes: scalar SHA-256
\texttt{7f5071e6...a539b759} or output-24 SHA-256
\texttt{e7bda332...5d2ca670}.  These are different models.  The Cube4
Transformer is another separate comparison.

\begin{table*}[t]
\centering
\small
\setlength{\tabcolsep}{3pt}
\caption{Native search inventory on one visible T4. This table records
algorithm type, compatible model family, and measured boundary; it is not a
cross-row speed ranking.}
\label{tab:four_t4}
\begin{tabular}{@{}p{3.0cm}p{2.2cm}p{3.4cm}p{2.5cm}p{3.0cm}p{2.2cm}@{}}
\toprule
Approach & Search type & Model family & Max complete & First failure/status & Valid metric \\
\midrule
AlphaCube 0.1.6 & beam & Cube3 action-18 & 1,572,864 & 1,703,936 SIGKILL & nodes/s \\
DeepCubeA & weighted A* & Cube3 scalar & n/a & complete, length 21 & iterations, nodes/s \\
Pilgrim Searcher & beam & Megaminx scalar & 5,898,240 & 6,291,456 OOM & wall/depth \\
Pilgrim QSearcher & beam & Megaminx output-24 & 20,971,520 & 23,068,672 OOM & wall/depth \\
native CayleyPy & beam & Megaminx scalar & 20,971,520 & targeted 21,037,056 OOM & wall/depth \\
MultiGPU-\newline BeamSearch & beam & Megaminx scalar & 29,360,128 & 31,457,280 OOM & wall/depth \\
MultiGPU-\newline BeamSearch & beam & Megaminx output-24 & 29,360,128 & 33,554,432 OOM & wall/depth \\
\bottomrule
\end{tabular}
\end{table*}

\paragraph{AlphaCube and DeepCubeA.}
For AlphaCube, s/depth is solver wall divided by the completed search depth
reported by the returned solution (20 for the first three rows and 17 for the
remaining rows).  For both Megaminx implementations it is wall divided by the
fixed eight completed depths.  These averages include startup and early
unsaturated layers; they complement, rather than replace, nodes/s.
AlphaCube materializes its $18B$ Cube3 children in host NumPy arrays. The
next tested width ended in SIGKILL, consistent with host-memory pressure,
but the released records do not confirm the kill cause with OOM diagnostics.
The largest
completed run reserved 4.686~GiB on the GPU; memory telemetry for the
failed run is not available in the released records.
Its widest completed rates are about 40--44 thousand nodes/s.  DeepCubeA is
weighted A*, not a persistent beam.  The audited run used weight $0.6$ and
search/NN batch sizes $10{,}000/10{,}000$; its T4 run completed 55 iterations
in 146.474 s, generated 6,134,352 nodes at 41,880 generated nodes/s, and found a length-21
solution.  These are A* statistics, not beam-depth or frontier-capacity values.
Figure~\ref{fig:alpha-curve} places the completed AlphaCube timings beside its
native node counter and identifies the next tested SIGKILL in the caption.

\begin{figure}[!ht]
\centering
\begin{tikzpicture}
\begin{groupplot}[
 group style={group size=1 by 2,vertical sep=1.25cm},
 width=0.90\columnwidth,height=4.3cm,
 xmode=log,log basis x=2,grid=both,
 tick label style={font=\scriptsize},label style={font=\small},
 scaled x ticks=false,
 xtick={1024,16384,262144,1048576},
 xticklabels={$2^{10}$,$2^{14}$,$2^{18}$,$2^{20}$}
]
\nextgroupplot[xlabel={beam width $B$},ylabel={mean s/depth},ymode=log]
\addplot+[thick,mark=*,blue!70!black] coordinates {
 (1024,.052) (4096,.060) (16384,.251) (65536,1.091)
 (131072,2.225) (262144,4.657) (524288,10.039)
 (786432,13.467) (1048576,17.875) (1310720,22.480)
 (1572864,27.146)
};
\nextgroupplot[xlabel={beam width $B$},ylabel={native nodes/s}]
\addplot+[thick,mark=square*,green!50!black] coordinates {
 (1024,18108) (4096,61309) (16384,56321) (65536,48722)
 (131072,46418) (262144,43698) (524288,40237)
 (786432,44222) (1048576,43670) (1310720,42955)
 (1572864,42388)
};
\end{groupplot}
\end{tikzpicture}
\caption{AlphaCube one-T4 completed sweep. The upper panel shows wall time per
completed depth; the lower uses AlphaCube's valid native node counter. The
next tested width, 1,703,936, ended in SIGKILL; host-memory pressure is an
inferred cause.}
\label{fig:alpha-curve}
\end{figure}

\paragraph{CayleyPy sweep.}
Figure~\ref{fig:cayleypy-curve} pairs the native CayleyPy timing curve with its
allocator growth; the failed broad-grid point is shown only as a memory
boundary, not as a completed timing.
\begin{figure}[!ht]
\centering
\begin{tikzpicture}
\begin{groupplot}[
 group style={group size=1 by 2,vertical sep=1.0cm},
 width=0.90\columnwidth,height=4.7cm,
 xmode=log,log basis x=2,grid=both,
 tick label style={font=\scriptsize},label style={font=\small},
 scaled x ticks=false,
 xtick={1024,16384,262144,4194304,33554432},
 xticklabels={$2^{10}$,$2^{14}$,$2^{18}$,$2^{22}$,$2^{25}$}
]
\nextgroupplot[xlabel={beam width $B$},ylabel={mean s/depth},ymode=log]
\addplot+[thick,mark=*,blue!70!black] coordinates {
 (1024,.116) (2048,.155) (4096,.274) (8192,.478)
 (16384,.847) (32768,1.715) (65536,3.925) (131072,7.958)
 (262144,14.319) (524288,27.499) (1048576,53.849)
 (2097152,93.168) (4194304,165) (8388608,311.169)
 (16777216,592.698)
};
\nextgroupplot[xlabel={beam width $B$},ylabel={allocator peak (MiB)},ymode=log,
 legend style={draw=none,font=\scriptsize,at={(0.03,0.97)},anchor=north west}]
\addplot+[thick,mark=*,green!50!black] coordinates {
 (1024,85.9) (2048,96.3) (4096,117.3) (8192,158.6)
 (16384,241.5) (32768,407.5) (65536,739.3) (131072,772.2)
 (262144,834.8) (524288,962.1) (1048576,1216.7)
 (2097152,1726) (4194304,2745.2) (8388608,5155.3)
 (16777216,10233.7)
};
\addlegendentry{allocated}
\addplot+[thick,mark=square*,orange!80!black] coordinates {
 (1024,92) (2048,112) (4096,128) (8192,230)
 (16384,292) (32768,510) (65536,858) (131072,1180)
 (262144,1168) (524288,2156) (1048576,3234)
 (2097152,2740) (4194304,4492) (8388608,6462)
 (16777216,14652)
};
\addlegendentry{reserved}
\addplot+[only marks,mark=x,mark size=4pt,very thick,black]
 coordinates {(33554432,14484)};
\addlegendentry{first broad-grid OOM}
\end{groupplot}
\end{tikzpicture}
\caption{Native CayleyPy one-T4 sweep. Time and allocator memory rise with
beam width; $2^{24}$ completes and $2^{25}$ is the first broad-grid OOM. The
failed point is not a timing measurement.}
\label{fig:cayleypy-curve}
\end{figure}

Pilgrim's widest scalar completion, $B=5{,}898{,}240$, took 1,462.557 s
(182.820 s/depth) with a 14,433 MiB device peak; $B=6{,}291{,}456$ CUDA-OOMed.
For output-24, $B=20{,}971{,}520$ completed in 197.358 s
(24.670 s/depth) at 12,333 MiB, whereas $B=23{,}068{,}672$ CUDA-OOMed. The
measured brackets are therefore 5,898,240--6,291,456 and
20,971,520--23,068,672, respectively.

The final collaborator-run MultiGPU pass and scalar-only v27 confirmation now
give measured complete/OOM pairs for both heads. Scalar completed at
29,360,128 and CUDA-OOMed at 31,457,280; native output-24 completed through
29,360,128 and CUDA-OOMed at 33,554,432.
Figure~\ref{fig:oom-brackets} juxtaposes those measured endpoints while keeping
the scalar and output-24 model families separate.

\begin{figure}[!ht]
\centering
\begin{tikzpicture}
\begin{groupplot}[
 group style={group size=1 by 2,vertical sep=1.0cm},
 width=0.90\columnwidth,height=4.4cm,
 xmode=log,log basis x=2,grid=major,
 tick label style={font=\scriptsize},label style={font=\small},
 scaled x ticks=false,
 xtick={4194304,8388608,16777216,33554432},
 xticklabels={$2^{22}$,$2^{23}$,$2^{24}$,$2^{25}$},
 ymin=0.5,ymax=3.5,
 legend style={draw=none,font=\scriptsize,at={(0.03,0.97)},anchor=north west}
]
\nextgroupplot[title={scalar model},
 ytick={1,2,3},yticklabels={Pilgrim,CayleyPy,MultiGPU}]
\addplot+[very thick,gray,mark=none,forget plot] coordinates {(5898240,1)(6291456,1)};
\addplot+[very thick,gray,mark=none,forget plot] coordinates {(20971520,2)(21037056,2)};
\addplot+[very thick,gray,mark=none,forget plot] coordinates {(29360128,3)(31457280,3)};
\addplot+[only marks,mark=*,blue!70!black,mark size=2.8pt]
 coordinates {(5898240,1)(20971520,2)(29360128,3)};
\addlegendentry{largest complete}
\addplot+[only marks,mark=x,black,very thick,mark size=3.5pt]
 coordinates {(6291456,1)(21037056,2)(31457280,3)};
\addlegendentry{first measured failure}
\nextgroupplot[xlabel={beam width $B$},title={output-24 model},
 ytick={1,2},yticklabels={Pilgrim,MultiGPU},ymin=0.5,ymax=2.5]
\addplot+[very thick,gray,mark=none,forget plot] coordinates {(20971520,1)(23068672,1)};
\addplot+[very thick,gray,mark=none,forget plot] coordinates {(29360128,2)(33554432,2)};
\addplot+[only marks,mark=*,blue!70!black,mark size=2.8pt]
 coordinates {(20971520,1)(29360128,2)};
\addplot+[only marks,mark=x,black,very thick,mark size=3.5pt]
 coordinates {(23068672,1)(33554432,2)};
\end{groupplot}
\end{tikzpicture}
\caption{Measured completion/failure brackets on one T4. A line joins the
largest completed width to the first measured failure; it does not claim that
the exact capacity lies at either endpoint. Scalar and output-24 models remain
separate.}
\label{fig:oom-brackets}
\end{figure}

\begin{table}[t]
\centering
\small
\caption{Final MultiGPU 1xT4 scalar boundary points.}
\label{tab:multigpu_v11_partial}
\resizebox{\columnwidth}{!}{%
\begin{tabular}{@{}rrrrl@{}}
\toprule
$B$ & Wall (s) & s/depth & Peak device (MiB) & Status \\
\midrule
25,165,824 & 3,027.673 & 378.459 & 12,267 & complete \\
27,262,976 & 3,080.206 & 385.026 & 13,241 & complete \\
29,360,128 & 2,817.158 & 352.145 & 14,215 & complete \\
31,457,280 & --- & --- & 14,909 & CUDA OOM \\
\bottomrule
\end{tabular}
}
\end{table}

The scalar OOM recurred for every row-budget profile from 49,152 through 24.
The common primary metric is total wall divided by eight, 352.145 s/depth;
the auxiliary final-depth time was 1,644.13 s.

\paragraph{MultiGPU output-24 sweep.}
The largest completion is $B=29{,}360{,}128$: 116.904 s total, 14.613 s/depth,
and 13,717 MiB peak device memory.  At $B=33{,}554{,}432$, every row-budget
profile down to 1 failed at 14,909 MiB.  Its final 1.563-s attempt is not a
completed depth time.  Figure~\ref{fig:multigpu_output24} therefore separates
the completed timing curve from the marked OOM boundary.

\begin{figure}[!ht]
\centering
\begin{tikzpicture}
\begin{groupplot}[
  group style={group size=1 by 2,vertical sep=1.0cm},
  width=0.88\columnwidth,height=4.4cm,
  xmode=log,log basis x=2,
  grid=both,
  tick label style={font=\scriptsize},
  label style={font=\small},
  scaled x ticks=false,
  xtick={1048576,4194304,8388608,16777216,33554432},
  xticklabels={$2^{20}$,$2^{22}$,$2^{23}$,$2^{24}$,$2^{25}$}
]
\nextgroupplot[xlabel={beam width $B$},ylabel={mean s/depth}]
\addplot+[thick,mark=*,blue!70!black] coordinates {
  (1048576,1.037) (4194304,2.842) (8388608,5.136)
  (16777216,9.683) (20971520,12.015) (25165824,13.891)
  (29360128,14.613)
};
\nextgroupplot[xlabel={beam width $B$},xlabel style={name=outputxlabel},
  ylabel={peak device memory (MiB)},
  legend style={font=\scriptsize,draw=none,at={(0.03,0.97)},anchor=north west}]
\addplot+[thick,mark=*,orange!80!black] coordinates {
  (1048576,773) (4194304,2253) (8388608,4159)
  (16777216,7983) (20971520,9893) (25165824,11799)
  (29360128,13717)
};
\addlegendentry{complete}
\addplot+[only marks,mark=x,mark size=4pt,very thick,black]
  coordinates {(33554432,14909)};
\addlegendentry{CUDA OOM}
\end{groupplot}
\node[draw,rounded corners,fill=blue!4,below=4mm of outputxlabel.south,
      text width=0.82\columnwidth,align=center,inner sep=2mm,font=\scriptsize]
  {\textbf{What happens when we double the frontier?}\\
   $B: 2^{22}\rightarrow2^{23}\rightarrow2^{24}$\\
   time: $2.842\rightarrow5.136\rightarrow9.683$ s/depth
   \quad($\times1.81,\ \times1.89$)\\
   memory: $2253\rightarrow4159\rightarrow7983$ MiB
   \quad($\times1.85,\ \times1.92$)};
\end{tikzpicture}
\caption{Native output-24 sweep on one T4. Upper: completed mean time per
depth. Lower: device-memory growth and the failed $B=33{,}554{,}432$ attempt.
The doubling rows are descriptive adjacent ratios from three completed points
in this same sweep, not a fit or asymptotic claim.  The OOM point is not treated
as a timing measurement.}
\label{fig:multigpu_output24}
\par\medskip
\noindent\begin{minipage}[t]{\columnwidth}
\paragraph{Reference-count comparison (not a performance fit).}
For fixed branching $a$, doubling retained $B$ doubles generated
$N=aB$ and frontier-state bytes.  In the sort-based abstract reference of
Eq.~\eqref{eq:reference-complexity}, the $N\log_2N$ reference-work term scales by
\begin{equation}
 \frac{(2N)\log_2(2N)}{N\log_2N}
 =2\!\left(1+\frac{1}{\log_2N}\right).
 \label{eq:doubling-reference-sort}
\end{equation}
These are operation and storage counts, not a wall-time bound for the
historical sweep or the audited Stream~4 dispatcher.  Fixed costs, occupancy,
other work, and repeated jobs can keep the measured adjacent ratios from
equaling two.
\end{minipage}
\end{figure}

\paragraph{Matched scalar cohorts.}
The matched scalar sweep completed CayleyPy through $2^{24}$ and OOMed at
$2^{25}$.  A separate targeted run gives the finer bracket
20,971,520-complete versus 21,037,056-OOM.  MultiGPU reached 29,360,128 at
14,215 MiB and OOMed at 31,457,280; its output-24 path reached 29,360,128 and
OOMed at 33,554,432.  Neither scalar path exposed a valid node counter, so only
wall time and memory are compared.  At $B=16{,}777{,}216$, wall time was
4,741.581 s for CayleyPy and 1,728.460 s for MultiGPU, a $2.74\times$ ratio of the recorded times under these timing boundaries.  Superseded and cancelled runs are excluded.

The broad matched curve is cohort C1 and uses the solver-internal MultiGPU
timer.  The later boundary rows are cohort C2: their executable-wrapper timer
includes initialization but excludes build and model export.  At
$B=25{,}165{,}824$, C1 records 2,416.150 s whereas C2 records 3,027.673 s.
Both are normalized by eight completed steps, but their timing boundaries
differ; the cohorts are neither pooled nor substituted for one another.
Figure~\ref{fig:megaminx_curve} therefore shows only the matched C1 curve and
states its timing boundary in the caption.

\begin{figure}[!ht]
\centering
\begin{tikzpicture}
\begin{axis}[
  width=0.96\columnwidth,height=5.5cm,
  xmode=log,log basis x=2,ymode=log,
  xlabel={retained beam width $B$},
  ylabel={mean wall time per depth (s)},
  grid=both,
  legend style={draw=none,font=\small,at={(0.03,0.97)},anchor=north west},
  tick label style={font=\scriptsize},
  label style={font=\small},
  scaled x ticks=false,
  xtick={65536,262144,1048576,4194304,16777216},
  xticklabels={$2^{16}$,$2^{18}$,$2^{20}$,$2^{22}$,$2^{24}$}
]
\addplot+[thick,mark=*,blue!70!black] coordinates {
  (65536,1.567) (262144,4.992) (1048576,19.076)
  (4194304,59.501) (8388608,111.619) (16777216,216.058)
  (20971520,272.499) (25165824,302.019)
};
\addlegendentry{MultiGPUBeamSearch}
\addplot+[thick,mark=square*,red!70!black] coordinates {
  (65536,3.925) (262144,14.319) (1048576,53.849)
  (4194304,165.000) (8388608,311.169) (16777216,592.698)
  (20971520,532.325)
};
\addlegendentry{CayleyPy}
\end{axis}
\end{tikzpicture}
\caption{Matched one-T4 Megaminx scalar sweep. Both axes are logarithmic.
The historical MultiGPU C1 series uses the solver-internal timer; CayleyPy uses
its recorded run wall. Values are normalized by eight completed steps. Neither
implementation exposes a valid node counter for these runs.}
\label{fig:megaminx_curve}
\end{figure}

Scalar timings are not compared with the faster, architecturally different
output-24 model.  In the separately matched Cube4 test, our tenth-layer pair
rate is $1.093\times$ Pilgrim on T4 and $1.042\times$ on P100.  The memory
counters differ (allocator reservation versus device-wide usage), so no memory
ratio is claimed.

The dated public-source audit located the same-metric chronology
\begin{equation}
55\longrightarrow54\longrightarrow53\longrightarrow51\ \text{moves}.
\label{eq:megaminx-chronology}
\end{equation}
The immediately preceding located public result is 53 moves.  The companion
record for the 51-move path identifies the same signed
72-degree, 24-generator metric and lists $22{,}600.2$ s
~\cite{megaminxsuperflip55,megaminxsuperflip54,megaminxsuperflip53,
megaminxsuperflip51}.  Independent CPU replay reaches the exact target.  The
audit located no shorter public same-metric path through August 25, 2026, but
there is no formal registry or optimality proof; we therefore do not call it a
world record.  This result is separate from the capacity and throughput runs.

\FloatBarrier
\section{Code and Artifacts}
\begin{table}[t]
\centering
\scriptsize
\caption{Public artifact map. Short link labels keep the paper readable; each
artifact is tied to the claim it can support.}
\label{tab:artifact-map}
\begin{tabular}{@{}p{2.2cm}p{4.8cm}@{}}
\toprule
Resource & Contents and supported claim \\
\midrule
\href{https://github.com/TryDotAtwo/MultiGPUBeamSearchPaper}{Paper repository}
& sources, evidence index, tables, notebooks; definitions and provenance \\
\href{https://github.com/TryDotAtwo/MultiGPUBeamSearchPaper/releases/tag/v1.0.0}{Paper release v1.0.0}
& matched checkpoints and PDFs; model-contract comparisons \\
\href{https://www.kaggle.com/datasets/trydotatwo/gpu-beam-search-paper-1xt4-reviewer-notebooks}{One-T4 dataset}
& data and notebooks; one-T4 sweeps and boundaries \\
\href{https://huggingface.co/datasets/TryDotAtwo/multigpu-beam-rtx3060-scaling-20261004}{RTX 3060 scaling dataset}
& 16 measured points, profiles, manifests and logs; matched 1/2/4/8-GPU scaling \\
\href{https://github.com/TryDotAtwo/MultiGPUBeamSearch}{Solver repository}
& solver, Megaminx release and replay audit; implementation and 51-move path \\
\bottomrule
\end{tabular}
\end{table}
{\small\raggedright
Evidence is revision-scoped.  The public index pins one-T4 to
\texttt{a1db0e6d9bb5458c8a842b37dfa99572d3025667}; Cube4 to
\texttt{df534ebb1ef624abd42443211778b5588916d6d2}; cluster/capacity to
\texttt{2d5f978fcfc1f018813fc49e9d9aeec32ea5e22c}; and CayleyPy to
\texttt{5b2b53f}.  It gives no measured solver revision for the two-T4
throughput record.  The H200 depth is pinned separately to
\texttt{3b756afe9e7636e53520fa6de7e60a167bbaa79a}.  None is relabeled as
\texttt{main@25cc25d}.  The RTX 3060 record is the public dataset observed at
revision \texttt{4f8e9b67fc00a2ef2e49ae1d2b255f860531d0aa}; its exact source
identities are listed in Appendix~\ref{app:rtx3060-scaling}.  In snapshot
\texttt{b113582e15c78d33deceea5922eb6589ae492fe4},
the actual SHA-256 of \texttt{artifacts/manifest.json} is
\texttt{c01dc1c3c74b78fc}\allowbreak
\texttt{8af4ba363f80edb5}\allowbreak
\texttt{f9aca4dbc10fb800}\allowbreak
\texttt{4b50ae38b209972c};
the older checksum list contains a different value and is not treated as a
verified inventory.\par}
{\small\raggedright
The complexity analysis is a separate read-only source audit, not a benchmark.
Its note has SHA-256
\texttt{98b8eea288b20b69}\allowbreak
\texttt{9a0bab4e962fc9f6}\allowbreak
\texttt{de6fc9ededa6698d}\allowbreak
\texttt{574ae3593d4e9528};
its source manifest has SHA-256
\texttt{a0f44e50a6aeb1b2}\allowbreak
\texttt{2fb0f48fa516c5d3}\allowbreak
\texttt{0b1b1c666c3c4b96}\allowbreak
\texttt{95aa43f6c94af70b}.
The manifest pins current dirty HEAD
\texttt{81ed9f0287c1f9057648119a253f208c31f7a872} and the exact SHA-256 of each
audited source file.  That snapshot, historical pin \texttt{a1db0e6}, and the
abstract exact-beam contract are three distinct scopes.  Current specialized
state widths and current final-union code are not substituted into historical
H200, T4, or A100 measurements.\par}
The A100 record supports capacity only; the T4 record supports Megaminx
throughput; the H200 record supports one completed Cube4 depth; and the RTX
3060 record supports only the declared profile-sensitive scaling series.
Missing fields are never borrowed from a different run.

\FloatBarrier
\section{Conclusion}
The conditional abstract beam-selection result can be summarized in one line:
\begin{equation}
\boxed{\frontier_{d,z}^{\rm distributed}(L)
=\frontier_{d,z}^{\rm monolithic}(L)}
\label{eq:conclusion}
\end{equation}
for the same complete candidate set, quantized scores, Hash128 key
equivalence, payload order, fully specified rank/physical-buffer/local-slot tie
order, distinct-key threshold witnesses, global one-per-key reduction,
race-free routing, and complete processing without semantic local caps.
The equality is selected membership under the same realized physical layout,
not a canonical serialization order or schedule-independent result from GPU
and shard counts alone.  Full-state equality additionally requires no relevant Hash128
collision.  This is beam-selection exactness, not shortest-path optimality.
The static audit leaves cross-buffer de-duplication unresolved and identifies a
multi-rank in-place owner-scatter risk in \texttt{a1db0e6}; neither finding is
evidence that its historical one-T4 measurement failed.  The H200 revision and
historical two-T4 \texttt{v118} run inherit neither the risk nor correctness
without their own code comparison.  Eight H200s completed one saturated
Cube4 depth at 2,900,361,216 retained records and a derived 74.746 million
nominal pairs/s; two T4s yielded a derived 30.274 million nominal
Megaminx parent--generator pairs/s;
and the older eight-A100 record establishes only requested capacity.  Those
cross-platform records do not establish strong scaling.  On the separate RTX
3060 host, fixed-count common/selected eight-GPU speedups were
$7.487/5.605$ (efficiencies $93.59\%/70.06\%$), while weak actual-work gains
were $6.113/7.295$ (efficiencies $76.42\%/91.19\%$).  The one-GPU medium
profile changed $85.107$ s to $63.595$ s while eight-GPU time stayed near
$11.35$ s: GPU count alone is therefore an incomplete performance label.
These data establish neither a causal bottleneck attribution nor identical
final search trajectories.  The artifacts separately contain a replay-valid
51-move Megaminx-superflip path, the shortest public same-metric solution
located by the dated audit.

\FloatBarrier
\section*{Acknowledgements}
The author thanks OpenAI, its AI tools including Codex, and the CayleyPy
project.
The author also thanks Vlad Kuznetsov for providing the neural network model
used in measurement runs and in the search that found the 51-move Megaminx
superflip path.

\section*{Use of Generative AI and Author Contributions}
OpenAI AI systems were used extensively in the development of
MultiGPUBeamSearch and the preparation of this paper. The author defined the
system architecture, directed the development and experimental work, and
personally checked the correctness of obtained solutions through Kaggle. The
author also personally verified the theorem relating threshold-based selection
to global top-$k$ selection. Most of the software implementation was generated
and revised using OpenAI Codex, which was also used for testing, debugging,
performance investigation, and experimental tooling. AI agents prepared
instrumentation and launch scripts, executed the RTX 3060 scaling experiments,
processed their recorded outputs, and generated tables and plots.

OpenAI assistants, including the assistant used in Prism, contributed to
mathematical analysis, theorem and proof drafting, complexity analysis,
literature and source checks, manuscript drafting and revision, and LaTeX
preparation. Numerical consistency checks, code-to-claim audits, and visual
checks of the manuscript were likewise AI-assisted. These checks have the
specific scope and limitations described in the paper; they do not establish
exhaustive implementation correctness or independent human verification of
every output. The author remains responsible for the submitted manuscript, its
claims, and its limitations.

\appendix
\section{Complexity Details for the Audited Code}
\label{app:complexity}
This appendix expands only the terms needed to interpret
Eqs.~\eqref{eq:stream3-complexity}--\eqref{eq:pipeline-time-model}.  It does not
assert that the current dirty snapshot produced any reported measurement or
satisfies the abstract exactness obligations.

\paragraph{Repeated reduction and final selection.}
Stream~3 comparison-sorts the full $Q$ capacity, including sentinel tail, then
reduces equal Hash128 keys.  Fixed-width radix operations used later for score
bins or shard indices do not make this Hash128 comparison sort linear.  Stream~4
similarly sorts the scheduled physical capacity $C_k$.  Current
\texttt{stream3.cu:863--868} can request a small-dirty job, including
\texttt{dirty=0}, when enough clean records remain; queue and running guards do
not prove an amortized job bound.  The current final A/B union additionally has
\begin{equation}
 W_{\rm union}=\sum_j O(C^{\rm union}_j\log C^{\rm union}_j),
 \label{eq:union-complexity}
\end{equation}
followed by histogram and boundary scans accounted as
$O(V_{\rm scan}+GH+EGJH)$ for $J$ logical shards and $E$ threshold updates.
This union term belongs only to the audited current code; it is not assigned to
historical executions.  A threshold may retain all candidates, so it provides
data-dependent savings but no worst-case survivor reduction.

\paragraph{Communication and skew.}
If $U_{\rm remote}\leq U_3\leq N_d$, candidate payload is
$m_{\rm meta}U_{\rm remote}=O(m_{\rm meta}N_d)$.  Final selected metadata and
reconstructed states add
$O(B_dm_{\rm meta}+B_ds_{\rm state})$ aggregate bytes and
$O(B_d\ell)$ materialization work.  Uniform independent ownership would give
expected load near $U_3/G$, but it is an assumption: one owner can receive
$\Theta(U_3)$.  Per-peer count and payload calls add peer-count and latency
terms; over $q$ exchange rounds the count operations are $\Theta(qG^2)$
aggregate.  Fabric bisection bandwidth, port bandwidth, collectives, zero-size
peer calls, and skew must therefore remain explicit in a wall-time model.

\paragraph{Peak memory.}
The audited current planner has the phase structure
\begin{equation}
 \begin{aligned}
 M^{\rm arena}_r={}&\operatorname{align}
   (s_{\rm state}F^{\rm cap}_r)+\operatorname{align}(M_{\rm solved})\\
 &+\operatorname{align}\!\left(\max_{x\in\mathcal P}M_x\right),
 \end{aligned}
 \label{eq:arena-peak}
\end{equation}
where $\mathcal P$ contains compatible stream, selection, materialization, and
other phases.  Stream memory includes score/hash rings, send/receive rings,
both physical survivor buffers, comparison-sort keys/values and temporary
storage, histograms, and spill capacity.  Process peak additionally includes
model weights, scorer activations/workspace, CUDA/NCCL contexts, graph
metadata, and allocator/runtime overhead.  Thus arbitrary survivors require
$O(N_dm_{\rm meta})$ aggregate persistent metadata and potentially that order
on one owner; $O(N_dm_{\rm meta}/G)$ requires balance.  The current audit's
specialized allocated state widths are not used to replace the separately
sourced 96-byte hypothetical full-child value in
Eq.~\eqref{eq:raw-memory} or any historical memory measurement.

\paragraph{History, depth, and wall time.}
Let $n_{{\rm hist},d}$ be final history entries recorded at depth $d$ and $h$
their stored bytes.  History work is
$\Theta(\sum_dn_{{\rm hist},d})$, device-to-host volume is
$m_{\rm meta}\sum_dn_{{\rm hist},d}$, and stored logical bytes are
$h\sum_dn_{{\rm hist},d}$; these counts are not all $N_d$ expansions.  With a
saturated unpruned beam this is $O(DBh)$.  Disk bandwidth and flush/read latency
remain service limits, and one reconstructed path can require $O(D)$ reads and
$O(D\ell)$ replay work.

For stage-work times $t_i$, network/history service time $t_{\rm service}$,
and final-drain time $t_{\rm final}$, assume completed bounded jobs and
no unaccounted scheduling or resource stalls. The scheduling model is
\begin{equation}
 \max\{t_i,t_{\rm service},t_{\rm final}\}\leq T_d
 \lesssim\sum_i t_i+t_{\rm service}+t_{\rm final}.
 \label{eq:time-bounds}
\end{equation}
The upper expression is serialized-work accounting; completion alone
does not bound unmeasured waiting or resource stalls.
The maximum form in Eq.~\eqref{eq:pipeline-time-model} requires demonstrated
overlap and independent resources.  Balance, bandwidth, latency, skew,
minimum job capacities, HBM contention, history service, and the final drain
bound strong scaling; no concrete crossover GPU count follows without measured
parameters.

\input{rtx3060_scaling_appendix}

\FloatBarrier
\begingroup
\footnotesize
\bibliographystyle{abbrv}
\bibliography{references}
\endgroup
\end{document}

%% file: rtx3060_scaling.tex
\subsection{Matched 1/2/4/8-GPU RTX 3060 Scaling}
\label{sec:rtx3060-scaling}
A separate public record measures one search on an exclusive host with eight
RTX 3060 12-GiB GPUs.  Every point uses the p900 input, the original FP16 MLP
with FP32 accumulation (not the Cube4 Transformer), seed
\texttt{0xC0DEC0DE}, and 12 complete steps.  The synchronized maximum-rank
time includes expansion, inference, exchange, global selection,
de-duplication, and history flush; initialization and model loading are
separate.  Each point had a full warmup and three timed repeats.

The strong series fixes $B=4{,}194{,}304$ and
$W=637{,}616{,}736$ expanded actions at every GPU count.  Its speedup and
efficiency are
\begin{equation}
 S_p=T_1/T_p,\qquad E^{\rm strong}_p=S_p/p,
 \label{eq:rtx3060-strong}
\end{equation}
using each profile series' own one-GPU median.  The selected profiles are
medium/common/common/common for $p=1/2/4/8$.  Medium changes the one-GPU
median from $85.107$ to $63.595$ s, while both eight-GPU medians remain near
$11.35$ s.  Consequently $S_8$ changes from $7.487$ to $5.605$ and efficiency
from $93.59\%$ to $70.06\%$.

The weak series uses $B_p=p\,4{,}194{,}304$ and expanded-action counts
$W_p=637{,}616{,}736$, $1{,}241{,}596{,}512$, $2{,}449{,}556{,}064$, and
$4{,}501{,}367{,}640$.  Every configuration starts from the same single state,
and its early frontier grows naturally, so $W_p$ is not exactly $pW_1$.  We
therefore report actual-work throughput gain rather than fixed-work speedup:
\begin{equation}
 G_p=\frac{W_p/T_p}{W_1/T_1},\qquad E^{\rm weak}_p=G_p/p.
 \label{eq:rtx3060-weak}
\end{equation}
Here $W_8/W_1=7.059676$.  The selected weak profile is large at every $p$;
common/selected $G_8$ is $6.113/7.295$, with efficiency
$76.42\%/91.19\%$.  Thus GPU count alone is not a complete performance label:
fixed beam and action counts still permit different batching, concurrency,
buffering, physical capacity, memory use, and executed system work.

\begin{table*}[!t]
\centering
\scriptsize
\caption{RTX 3060 scaling medians.  Times are rounded to three decimals here;
the public CSV contains exact summaries and paths to all three repeats.  $R$ denotes
$S_p$ for strong rows and actual-work $G_p$ for weak rows.  The daggered weak
common $p=2$ point is retained but its $5.97\%$ min--max range/median exceeds
the $5\%$ winner-selection stability gate.}
\label{tab:rtx3060-scaling}
\setlength{\tabcolsep}{4pt}
\begin{tabular}{@{}llr rr l rr@{}}
\toprule
Regime & $p$ & Expanded actions $W_p$ & Common $T_p$ (s) & Common $R$
& Selected profile & Selected $T_p$ (s) & Selected $R$ \\
\midrule
Strong & 1 & 637,616,736 & 85.107 & 1.000 & medium & 63.595 & 1.000 \\
Strong & 2 & 637,616,736 & 33.952 & 2.507 & common & 33.998 & 1.871 \\
Strong & 4 & 637,616,736 & 18.727 & 4.545 & common & 18.789 & 3.385 \\
Strong & 8 & 637,616,736 & 11.367 & 7.487 & common & 11.347 & 5.605 \\
\midrule
Weak & 1 & 637,616,736 & 100.779 & 1.000 & large & 72.209 & 1.000 \\
Weak & 2 & 1,241,596,512 & $117.355^{\dagger}$ & 1.672 & large & 69.980 & 2.009 \\
Weak & 4 & 2,449,556,064 & 120.793 & 3.205 & large & 70.951 & 3.910 \\
Weak & 8 & 4,501,367,640 & 116.380 & 6.113 & large & 69.876 & 7.295 \\
\bottomrule
\end{tabular}
\end{table*}

\begin{figure*}[!t]
\centering
\begin{tikzpicture}
\begin{groupplot}[
  group style={group size=2 by 2,horizontal sep=1.25cm,vertical sep=1.30cm},
  width=0.42\textwidth,
  height=0.245\textwidth,
  xmin=0.5,xmax=8.5,
  xtick={1,2,4,8},
  xlabel={GPUs $p$},
  grid=major,
  tick label style={font=\scriptsize},
  label style={font=\scriptsize},
  title style={font=\small\bfseries},
  legend style={font=\scriptsize,legend columns=2,draw=none,
                /tikz/every even column/.append style={column sep=7mm}},
  error bars/error bar style={line width=0.55pt},
  error bars/error mark options={rotate=90,mark size=2pt}
]
\nextgroupplot[
  title={Strong time: fixed $B,W$},
  xlabel={},
  ylabel={whole-search median (s)},
  ymin=0,ymax=95,
  legend to name=rtx3060legend
]
\addplot+[blue!70!black,thick,mark=*,
  error bars/.cd,y dir=both,y explicit]
  table[x=p,y=t,y error minus=em,y error plus=ep] {
p t em ep
1 85.107276547 0.073228518 0.000173233
2 33.951948256 0.051202789 0.026273500
4 18.727284617 0.034115685 0.010739788
8 11.366969677 0.000766408 0.079189481
};
\addlegendentry{common}
\addplot+[orange!85!black,thick,mark=square*,
  error bars/.cd,y dir=both,y explicit]
  table[x=p,y=t,y error minus=em,y error plus=ep] {
p t em ep
1 63.595126667 0.053387093 0.028622512
2 33.998143934 0.042552172 0.017765227
4 18.789272904 0.047873015 0.001180460
8 11.346553443 0.008253718 0.032783712
};
\addlegendentry{selected}
\addplot[blue!70!black,densely dashed,mark=none,forget plot]
  coordinates {(1,85.107276547) (2,42.553638273) (4,21.276819137) (8,10.638409568)};
\addplot[orange!85!black,densely dashed,mark=none,forget plot]
  coordinates {(1,63.595126667) (2,31.797563334) (4,15.898781667) (8,7.949390833)};

\nextgroupplot[
  title={Strong own-baseline speedup},
  xlabel={},
  ylabel={$S_p=T_1/T_p$},
  ymin=0,ymax=8.5
]
\addplot+[blue!70!black,thick,mark=*,
  error bars/.cd,y dir=both,y explicit]
  table[x=p,y=g,y error minus=em,y error plus=ep] {
p g em ep
1 1.000000000 0.000002035 0.000861167
2 2.506697875 0.001938292 0.003786050
4 4.544560425 0.002604736 0.008293981
8 7.487244091 0.051799994 0.000504855
};
\addplot+[orange!85!black,thick,mark=square*,
  error bars/.cd,y dir=both,y explicit]
  table[x=p,y=g,y error minus=em,y error plus=ep] {
p g em ep
1 1.000000000 0.000449872 0.000840189
2 1.870547015 0.000976916 0.002344116
4 3.384650752 0.000212632 0.008645749
8 5.604796821 0.016147342 0.004080013
};
\addplot[black,densely dashed,mark=none,forget plot]
  coordinates {(1,1) (2,2) (4,4) (8,8)};

\nextgroupplot[
  title={Weak time: growing $B_p,W_p$},
  ylabel={whole-search median (s)},
  ymin=55,ymax=130
]
\addplot+[blue!70!black,thick,mark=*,
  error bars/.cd,y dir=both,y explicit]
  table[x=p,y=t,y error minus=em,y error plus=ep] {
p t em ep
1 100.779434612 0.062609872 0.085644902
2 117.355044751 4.937967338 2.069997647
4 120.793078979 2.440427145 3.019695625
8 116.380041126 0.119730427 0.382858980
};
\addplot+[orange!85!black,thick,mark=square*,
  error bars/.cd,y dir=both,y explicit]
  table[x=p,y=t,y error minus=em,y error plus=ep] {
p t em ep
1 72.209107991 0.176919254 0.019386920
2 69.979927514 0.023945464 0.003713395
4 70.951101770 0.155651067 0.071967694
8 69.875745411 0.102579733 0.225378658
};
\addplot[blue!70!black,dotted,mark=none,forget plot]
  coordinates {(1,100.779434612) (2,100.779434612) (4,100.779434612) (8,100.779434612)};
\addplot[orange!85!black,dotted,mark=none,forget plot]
  coordinates {(1,72.209107991) (2,72.209107991) (4,72.209107991) (8,72.209107991)};
\addplot[blue!70!black,densely dashed,mark=none,forget plot]
  coordinates {(1,100.779434612) (2,98.121165452) (4,96.792030871) (8,88.933770894)};
\addplot[orange!85!black,densely dashed,mark=none,forget plot]
  coordinates {(1,72.209107991) (2,70.304441175) (4,69.352107767) (8,63.721614348)};

\nextgroupplot[
  title={Weak actual-work throughput gain},
  ylabel={$G_p=(W_p/T_p)/(W_1/T_1)$},
  ymin=0,ymax=8.5
]
\addplot+[blue!70!black,thick,mark=*,
  error bars/.cd,y dir=both,y explicit]
  table[x=p,y=g,y error minus=em,y error plus=ep] {
p g em ep
1 1.000000000 0.000849104 0.000621643
2 1.672210439 0.028984471 0.073452546
4 3.205217772 0.078172726 0.066091468
8 6.113334901 0.020045281 0.006295805
};
\addplot+[orange!85!black,thick,mark=square*,
  error bars/.cd,y dir=both,y explicit]
  table[x=p,y=g,y error minus=em,y error plus=ep] {
p g em ep
1 1.000000000 0.000268411 0.002456114
2 2.009274478 0.000106614 0.000687761
4 3.909853746 0.003961856 0.008596215
8 7.295420060 0.023455144 0.010725646
};
\addplot[black,densely dashed,mark=none,forget plot]
  coordinates {(1,1) (2,2) (4,4) (8,8)};
\end{groupplot}
\node[anchor=north] at
  ($(group c1r2.south)!0.5!(group c2r2.south)+(0,-8mm)$)
  {\pgfplotslegendfromname{rtx3060legend}};
\end{tikzpicture}
\caption{Profile-sensitive RTX 3060 scaling.  Solid lines are measured
three-repeat medians; bars span the observed minimum and maximum, not a
confidence interval.  Gain bars hold the corresponding $T_1$ median fixed.
In the strong-time panel, colored dashed lines are each series' $T_1/p$.
In the weak-time panel, dotted lines are constant $T_1$ and colored dashed
lines are $T_1(W_p/W_1)/p$, the ideal time for linear throughput on the actual
measured work.  Black dashed gain references are $y=p$.  No line is a fitted
performance law.}
\label{fig:rtx3060-scaling}
\end{figure*}

The common strong series is superlinear at two and four GPUs, but the
experiment does not isolate a cause; locality and single-versus-distributed
path explanations remain hypotheses.  Selected means fastest among the admitted stable candidates from three
tested profiles, followed by independent final repeats, not a global
optimum or an optimization ablation.  The first five strong frontier
memberships match across $p$; the first saturated cutoff differences are tied
at score 43,768 under rank/physical-buffer/local-slot order, so later
trajectories may differ.  The experimenter reports all 12 saved frontiers
checked for uniqueness, balance, and padding admission, plus fixed-$p$ weak
profile agreement.  These are bounded recorded checks, not an exhaustive
implementation proof or evidence of identical final search results.

%% file: rtx3060_scaling_appendix.tex
\section{RTX 3060 Scaling Protocol Details}
\label{app:rtx3060-scaling}
The three tested profiles change physical execution parameters, not the beam
semantics.  Every profile uses eight logical shards and four active sort slots.
Strong history allocation is 8 GiB total; weak allocation is 8 GiB per GPU,
with no history pruning.  This is capacity allocation, not a semantic shard
cap.

\begin{table*}[!t]
\centering
\scriptsize
\caption{RTX 3060 profile parameters.  Weak runs override only the physical
shard-buffer capacity to 3,276,800 for every profile and GPU count.  ``S4'' is
both batch and trigger capacity; the final materialization chunk has the same
size.}
\label{tab:rtx3060-profiles}
\setlength{\tabcolsep}{4pt}
\begin{tabular}{@{}lrrrrrrrr@{}}
\toprule
Profile & examples/forward & outer/lane & concurrency & ring & S3 accumulator
& strong shard cap & weak shard cap & S4/final chunk \\
\midrule
common & 8,192 & 196,608 & 1 & 2 & 393,216 & 655,360 & 3,276,800 & 65,536 \\
medium & 16,384 & 393,216 & 2 & 2 & 786,432 & 1,310,720 & 3,276,800 & 131,072 \\
large & 32,768 & 786,432 & 2 & 4 & 3,145,728 & 3,276,800 & 3,276,800 & 262,144 \\
\bottomrule
\end{tabular}
\end{table*}

\paragraph{Capacity amendment and selection.}
The original weak common $p=2$ admission with capacity 655,360 overflowed a
physical Stream~3 buffer before completion and is excluded from timing.  All
weak candidates were then admitted with capacity 3,276,800; weak $p=1$ was
remeasured independently rather than reused from strong scaling.  The runner,
global beams, 12-step workload, and global selection/de-duplication semantics
were unchanged.  The weak one-GPU common point ($100.779$ s) and strong
one-GPU common point ($85.107$ s) use different physical buffer capacities,
so their difference does not isolate a causal capacity effect.  Each candidate received a warmup and three selection
repeats; a winner-eligible candidate required range/median at most $5\%$.  The
frozen winner then received an independent warmup and three final repeats.
The reported weak common $p=2$ point has a $5.97\%$ range and therefore does
not satisfy that winner-selection gate.

\paragraph{Hardware and telemetry.}
The exclusive host had eight RTX 3060 12-GiB GPUs, two Xeon E5-2673 v4 CPUs,
and no NVLink; links were PIX/PHB within groups and NODE between groups.  The
record reports driver 595.71.05, CUDA compiler 12.8.93, GCC 13.3.0, NCCL
2.31.2, and Linux 6.8.0-124-generic.  GPU clocks were observed rather than
locked.  Device memory was sampled through \texttt{nvidia-smi} at roughly
0.2-s intervals plus command overhead, so transient peaks may be missed.
Min--max intervals are three-repeat ranges, not confidence intervals.  The
experiment contains no isolated capacity-effect or optimization ablation.

\paragraph{Measurement identities.}
The public dataset is
\path{TryDotAtwo/multigpu-beam-rtx3060-scaling-20261004},
observed revision
\texttt{4f8e9b67fc00a2ef2e49}\allowbreak
\texttt{ae1d2b255f860531d0aa}.  The final report SHA-256 is
\texttt{f87d0ef6d92115fd}\allowbreak
\texttt{bb1b83f05a92afce}\allowbreak
\texttt{d7d9fbb697663968}\allowbreak
\texttt{7867a53bad79b207};
the measurement CSV is
\texttt{926a3103df656ef4}\allowbreak
\texttt{27d89f62b51d7779}\allowbreak
\texttt{b965f8af1f61b0f2}\allowbreak
\texttt{3453735e787b3449};
and the completion audit is
\texttt{fc9112a534d9bed4}\allowbreak
\texttt{a5c6d704a26192de}\allowbreak
\texttt{8d6edc2f1a84139e}\allowbreak
\texttt{47d7ba640ce1d544}.
The immutable runner binary is
\texttt{7c10c4e8cade66d7}\allowbreak
\texttt{8900946052e0a1cc}\allowbreak
\texttt{f08644fa0df37df6}\allowbreak
\texttt{b5286fd3028279a2};
the instrumented runner is
\texttt{a38556e3fd5a8531}\allowbreak
\texttt{3d74b5fa1a573be1}\allowbreak
\texttt{bf26baa2586fb61c}\allowbreak
\texttt{a081d01327c7366b}.
The completion audit reverified 346 per-file source-manifest entries based on
\texttt{81ed9f0287c1f9057648119a253f208c31f7a872}.  Sharing that base revision
does not prove identity with the separately audited dirty snapshot; neither
correctness nor audit findings transfer without comparison of the per-file
bytes.

\paragraph{Admission scope.}
The experimenter reports all 12 saved frontiers checked for logical-state
uniqueness, distribution balance, and valid padding.  Strong memberships match
across $p$ for the first five frontiers; first saturated-cutoff differences
are score-43,768 ties under rank/physical-buffer/local-slot order.  At each
fixed weak $p$, all 12 saved frontiers match across tested profiles.  These are
bounded recorded admission checks.  They are not an independent exhaustive
implementation proof, a guarantee of identical later strong trajectories, or
evidence of identical final solutions.

%% file: references.bib
@inproceedings{zhou2015gpuastar,
  author = {Yichao Zhou and Jianyang Zeng},
  title = {Massively Parallel {A*} Search on a {GPU}},
  booktitle = {Proceedings of the AAAI Conference on Artificial Intelligence},
  year = {2015},
  doi = {10.1609/aaai.v29i1.9367}
}

@article{kishimoto2013hda,
  author = {Akihiro Kishimoto and Alex Fukunaga and Adi Botea},
  title = {Evaluation of a Simple, Scalable, Parallel Best-First Search Strategy},
  journal = {Artificial Intelligence},
  volume = {195},
  pages = {222--248},
  year = {2013},
  doi = {10.1016/j.artint.2012.10.007}
}

@article{chervov2025large,
  author = {Alexander Chervov and others},
  title = {A Machine Learning Approach That Beats Large Rubik's Cubes},
  journal = {arXiv preprint arXiv:2502.13266},
  year = {2025},
  url = {https://arxiv.org/abs/2502.13266}
}

@article{soibelman2025cayleypy,
  author = {Alexander Chervov and others},
  title = {{CayleyPy RL}: Pathfinding and Reinforcement Learning on Cayley Graphs},
  journal = {arXiv preprint arXiv:2502.18663},
  year = {2025},
  url = {https://arxiv.org/abs/2502.18663}
}

@article{he2021multigpuastar,
  author = {Xin He and Yapeng Yao and Zhiwen Chen and Jianhua Sun and Hao Chen},
  title = {Efficient Parallel {A*} Search on Multi-{GPU} System},
  journal = {Future Generation Computer Systems},
  volume = {123},
  pages = {35--47},
  year = {2021},
  doi = {10.1016/j.future.2021.04.011}
}

@inproceedings{kim2025pathweaver,
  author = {Sukjin Kim and Seongyeon Park and Si Ung Noh and Junguk Hong and Taehee Kwon and Hunseong Lim and Jinho Lee},
  title = {{PathWeaver}: A High-Throughput Multi-GPU System for Graph-Based Approximate Nearest Neighbor Search},
  booktitle = {2025 USENIX Annual Technical Conference},
  pages = {1501--1517},
  year = {2025},
  url = {https://www.usenix.org/conference/atc25/presentation/kim}
}

@inproceedings{anantharaman1986accelerator,
  author = {Thomas S. Anantharaman and Roberto Bisiani},
  title = {A Hardware Accelerator for Speech Recognition Algorithms},
  booktitle = {Proceedings of the 13th Annual International Symposium on Computer Architecture},
  pages = {216--223},
  year = {1986},
  doi = {10.1145/17356.17382}
}

@inproceedings{wijs2007distributed,
  author = {Anton Wijs and Bert Lisser},
  title = {Distributed Extended Beam Search for Quantitative Model Checking},
  booktitle = {Model Checking and Artificial Intelligence},
  pages = {166--184},
  year = {2007},
  doi = {10.1007/978-3-540-74128-2_11}
}

@techreport{chong2008lvcsr,
  author = {Jike Chong and Youngmin Yi and Arlo Faria and Nadathur Rajagopalan Satish and Kurt Keutzer},
  title = {Data-Parallel Large Vocabulary Continuous Speech Recognition on Graphics Processors},
  institution = {University of California, Berkeley},
  number = {UCB/EECS-2008-69},
  year = {2008},
  url = {https://www2.eecs.berkeley.edu/Pubs/TechRpts/2008/EECS-2008-69.html}
}

@inproceedings{frohner2022parallel,
  author = {Nikolaus Frohner and Jan Gmys and Nouredine Melab and G{\"u}nther R. Raidl and El-Ghazali Talbi},
  title = {Parallel Beam Search for Combinatorial Optimization},
  booktitle = {Proceedings of the 51st International Conference on Parallel Processing Workshops},
  year = {2022},
  doi = {10.1145/3547276.3548633}
}

@article{takano2023alphacube,
  author = {Kyo Takano},
  title = {Self-Supervision is All You Need for Solving Rubik's Cube},
  journal = {Transactions on Machine Learning Research},
  year = {2023},
  url = {https://openreview.net/forum?id=bnBeNFB27b}
}

@article{agostinelli2019deepcubea,
  author = {Forest Agostinelli and Stephen McAleer and Alexander Shmakov and Pierre Baldi},
  title = {Solving the Rubik's Cube with Deep Reinforcement Learning and Search},
  journal = {Nature Machine Intelligence},
  volume = {1},
  pages = {356--363},
  year = {2019},
  doi = {10.1038/s42256-019-0070-z}
}

@inproceedings{kuroiwa2024parallel,
  author = {Ryo Kuroiwa and J. Christopher Beck},
  title = {Parallel Beam Search Algorithms for Domain-Independent Dynamic Programming},
  booktitle = {Proceedings of the AAAI Conference on Artificial Intelligence},
  volume = {38},
  pages = {20743--20750},
  year = {2024},
  doi = {10.1609/aaai.v38i18.30062}
}

@article{wijs2024gpuexplore,
  author = {Anton Wijs and Muhammad Osama},
  title = {The Fast and the Capacious: Memory-Efficient Multi-{GPU} Accelerated Explicit State Space Exploration with {GPUexplore} 3.0},
  journal = {Frontiers in High Performance Computing},
  volume = {2},
  year = {2024},
  doi = {10.3389/fhpcp.2024.1285349}
}

@inproceedings{sun2026cunnqssci,
  author = {Daran Sun and Bowen Kan and Haoquan Long and Hairui Zhao and Haoxu Li and Yicheng Liu and Pengyu Zhou and Ankang Feng and Wenjing Huang and Yida Gu and Zhenyu Li and Honghui Shang and Yunquan Zhang and Dingwen Tao and Ninghui Sun and Guangming Tan},
  title = {A Fully {GPU}-Accelerated Framework for High-Performance Configuration Interaction Selection with Neural Network Quantum States},
  booktitle = {Proceedings of the 35th International Symposium on High-Performance Parallel and Distributed Computing},
  year = {2026},
  doi = {10.1145/3806645.3807583}
}

@inproceedings{soltani2026gpuastar,
  author = {Misagh Soltani and Forest Agostinelli},
  title = {{GPU}-Accelerated {A*} Search with Deep Neural Network Heuristics},
  booktitle = {Proceedings of the Nineteenth International Symposium on Combinatorial Search},
  year = {2026},
  url = {https://ojs.aaai.org/index.php/SOCS/article/download/43083/50643/47188}
}

@inproceedings{chan2025trie,
  author = {Brian J Chan and others},
  title = {Efficient Beam Search for Large Language Models Using Trie-Based Decoding},
  booktitle = {Proceedings of the 2025 Conference on Empirical Methods in Natural Language Processing},
  pages = {14795--14807},
  year = {2025},
  url = {https://aclanthology.org/2025.emnlp-main.748/}
}

@inproceedings{junczysdowmunt2016marian,
  author = {Marcin Junczys-Dowmunt and Tomasz Dwojak and Hieu Hoang},
  title = {Is Neural Machine Translation Ready for Deployment? A Case Study on 30 Translation Directions},
  booktitle = {Proceedings of the 13th International Conference on Spoken Language Translation},
  year = {2016},
  url = {https://aclanthology.org/2016.iwslt-1.5/}
}

@inproceedings{braun2020gpuasr,
  author = {Hugo Braun and Justin Luitjens and Ryan Leary and Tim Kaldewey and Daniel Povey},
  title = {{GPU}-Accelerated Viterbi Exact Lattice Decoder for Batched Online and Offline Speech Recognition},
  booktitle = {2020 IEEE International Conference on Acoustics, Speech and Signal Processing},
  year = {2020},
  doi = {10.1109/ICASSP40776.2020.9054099},
  url = {https://arxiv.org/abs/1910.10032}
}

@misc{megaminxsuperflip53,
  author = {{SpeedSolving forum contributor}},
  title = {Development of a Megaminx Solver: 53-Move Superflip Report},
  year = {2026},
  howpublished = {SpeedSolving forum post 1700151},
  url = {https://www.speedsolving.com/threads/development-of-a-megaminx-solver.93202/post-1700151}
}

@misc{megaminxsuperflip55,
  author = {{SpeedSolving forum contributor}},
  title = {Development of a Megaminx Solver: 55-Move Superflip Report},
  year = {2026},
  howpublished = {SpeedSolving forum post 1696242},
  url = {https://www.speedsolving.com/threads/development-of-a-megaminx-solver.93202/post-1696242}
}

@misc{megaminxsuperflip54,
  author = {{SpeedSolving forum contributor}},
  title = {Development of a Megaminx Solver: 54-Move Superflip Report},
  year = {2026},
  howpublished = {SpeedSolving forum post 1696217},
  url = {https://www.speedsolving.com/threads/development-of-a-megaminx-solver.93202/post-1696217}
}

@misc{megaminxsuperflip51,
  author = {{SpeedSolving forum contributor}},
  title = {Development of a Megaminx Solver: 51-Move Superflip Report},
  year = {2026},
  howpublished = {SpeedSolving forum post 1705451},
  url = {https://www.speedsolving.com/threads/development-of-a-megaminx-solver.93202/post-1705451}
}
